\documentclass[10pt,journal,compsoc]{IEEEtran}

\ifCLASSOPTIONcompsoc
  \usepackage[nocompress]{cite}
\else
 \usepackage{cite}
\fi

\usepackage{graphicx}
\usepackage{amssymb}
\usepackage{amsmath}

\usepackage{mathrsfs}
\usepackage[usenames, dvipsnames]{color}
\usepackage{float}
\usepackage{subfigure}
\usepackage{epstopdf}
\usepackage{mathrsfs}
\usepackage{caption}

\usepackage{lipsum}
\usepackage{algorithmic}

\newtheorem{theorem}{Theorem}

\newtheorem{proposition}{Proposition}

\newtheorem{example}{Example}

\newtheorem{remark}{Remark}
\newcommand{\argmax}{\operatornamewithlimits{argmax}}
\newcommand{\argmin}{\operatornamewithlimits{argmin}}

\newcommand{\ignore}[1]{}

\newcommand\boldblue[1]{\textcolor{blue}{\textbf{#1}}}

\begin{document}

\renewcommand\thepage{}
\title{\LARGE \bf Beyond the VCG Mechanism: Truthful Reverse Auctions for Relay Selection with High Data Rates, High Base Station Utility and Low Interference in D2D Networks}
\author{Aditya MVS, Harsh Pancholi, Priyanka P. and Gaurav S. Kasbekar

\IEEEcompsocitemizethanks{\IEEEcompsocthanksitem Aditya MVS and G. S. Kasbekar are with the Department of Electrical Engineering, Indian Institute of Technology Bombay, Mumbai, India. Their email addresses are adityamvs@iitb.ac.in and gskasbekar@ee.iitb.ac.in respectively. H. Pancholi is with Samsung R$\&$D Bangalore, India. His email address is h.pancholi@samsung.com. Priyanka P. is with Qualcomm Hyderabad, India. Her email address is ppriyank@qti.qualcomm.com.}

\IEEEcompsocitemizethanks{\IEEEcompsocthanksitem A preliminary version of this paper appeared in Proc. of NCC 2017~\cite{RF:NCC:2017}. }}

\IEEEtitleabstractindextext{
\begin{abstract}
Device-to-Device communication allows a cellular user (relay node) to
relay data between the base station (BS) and another cellular user
(destination node). We address the problem of designing reverse auctions
to assign a relay node to each destination node, when there are multiple
potential relay nodes and multiple destination nodes, in the scenarios
where the transmission powers of the relay nodes are: 1) fixed, 2)
selected to achieve the data rates desired by destination nodes, and 3)
selected so as to approximately maximize the BS's utility. We show that
auctions based on the widely used Vickrey-Clarke-Groves (VCG) mechanism
have several limitations in scenarios 1) and 2); also, in scenario 3), the
VCG mechanism is not applicable. Hence, we propose novel reverse auctions
for relay selection in each of the above three scenarios. We prove that
all the proposed reverse auctions can be truthfully implemented as well as
satisfy the individual rationality property. Using numerical computations,
we show that in scenarios 1) and 2), our proposed auctions significantly
outperform the auctions based on the VCG mechanism in terms of the data
rates achieved by destination nodes, utility of the BS and/ or the
interference cost incurred to the BS.
\end{abstract}
}

\maketitle

\IEEEraisesectionheading{\section{Introduction}\label{sec:introduction}}

The demand from mobile users is rapidly increasing due to the proliferation of new applications such as video streaming services, online gaming etc. Long-Term Evolution (LTE)-Advanced~\cite{RF:5G} is being extensively deployed worldwide, and 5G cellular networks are being researched upon~\cite{Andrews20141065}, to meet the growing demand. Some of the objectives of LTE-Advanced and 5G cellular networks   are to provide improved cell-edge capacity relative to LTE~\cite{RF:UTRA} and decreased consumption of energy. Issues such as low signal to noise ratio (mainly at the cell edges) and coverage holes due to shadowing have to be tackled for achieving these objectives. As the link capacity of current technology is already close to the Shannon bound~\cite{RF:Mogensen}, the deployment of additional network infrastructure such as low-power base stations and dedicated relay nodes is considered as a possible solution. However, this involves huge deployment costs. One alternative to avoid this is to use the concept of \emph{Device-to-Device (D2D) communication}~\cite{RF:D2D:survey}. D2D communication enables a  mobile device to directly communicate with its peers bypassing the base station (BS)~\cite{RF:doppler}. In this paper, we study a scenario where the BS requests some of the existing cellular users to act as relays between the BS and other cellular users to improve the throughput of cell-edge users and users that experience poor signal to noise ratio from the BS due to shadowing, and to extend the network coverage, i.e., the BS employs relaying using D2D communication.  This also replaces a single high-powered link with two low-powered links, which can increase the energy efficiency of the network.

D2D communications, an innovative technique for next generation cellular networks, makes the relaying concept simpler with no need of introducing extra relay nodes in the network~\cite{RF:D2D:survey}. Also, it was shown in~\cite{RF:Wen} that the achieved channel capacity in cellular networks in which D2D communication is used for relaying is enhanced when compared to the case without such relaying. We consider a scenario where D2D communication occurs \emph{underlay}, i.e., D2D communication takes place on the same set of channels as traditional cellular communication (communication between the BS and cellular users)~\cite{RF:D2D:survey}. Note that underlay D2D communication increases the interference caused to the traditional cellular communication users. However, it is shown in~\cite{RF:chia-hao} that through proper sharing of resources between the tradional cellular communication users and D2D users and control of transmission power, underlay D2D communication increases the overall throughput of the network. 

As relaying of data (to another user with poor channel conditions from the BS) consumes energy, cellular users may not be willing to relay, since they would want to conserve battery energy for personal use in future. Thus, \emph{incentives} must be provided by the centralized entity (BS or eNodeB) to make potential relays cooperate for throughput enhancement.  In addition, although a BS can increase the achieved data rate of its cellular user experiencing poor channel conditions from the BS by selecting a relay which is willing to forward data to it, this will also increase the interference caused by the relay to its traditional cellular communication user which is using the same channel.  So the costs incurred to the BS are: the incentives provided to the relay and the interference caused by the relay to its traditional cellular communication user. Thus, a BS has to select relays which can increase the throughput of the users experiencing poor channel conditions from the BS at a minimum expense to the BS and minimum interference to its traditional cellular communication users.

Apart from normal relaying, in which first the BS sends the message to the relay, which is ignored by the destination node, and then the relay forwards the message to the destination node, different \emph{cooperative relaying schemes}~\cite{RF:Laneman} such as amplify-and-forward, decode-and-forward and selection relaying can be used. In each of the latter three schemes, the BS (source) transmits the message in the first time slot. Both the destination node and the relay receive this transmission in the first time slot. The relay node processes the received message (depending on the relaying scheme used) and sends it to the destination node in the second time slot. The destination node combines the transmission by the source in the first slot and by the relay in the second slot to form the received message. Cooperative relaying schemes have the advantage that they exploit space diversity to improve the achieved data rate~\cite{RF:Laneman}. 

In this paper, we consider a BS which requires relays to communicate with some of its cellular users (henceforth referred to as \emph{destination nodes}) when the BS cannot communicate with the latter directly at sufficiently high data rates due to network coverage problems. We study the problem of designing a \emph{reverse auction} conducted by the BS in which the BS requests some of its users (henceforth referred to as \emph{relay nodes}) to act as relays to its destination nodes. The BS must provide incentives (e.g., monetary payments) to the relay nodes for acting as relays, since relays incur a cost due to their battery drain. For an auction to be feasible, the incentive provided by the BS to a relay node must be at least the cost incurred by the node for acting as a relay or else no node will participate in the auction. However, a relay node's incurred cost is private information of the node and is not known to the BS. Thus a greedy relay node can falsely declare the cost it incurs. Hence, mechanisms are required for ensuring that relay nodes \emph{truthfully} declare the costs they incur. In this paper, we address the problem of designing reverse auctions that induce relay nodes to truthfully declare their costs for three different scenarios: 1) Constant power case, where the BS assigns a fixed transmission power to all the relay nodes, 2) Constant data rate case, where each destination node requests the BS for a desired data rate and the relay node assigned to a destination node must transmit at a power such that the desired data rate is achieved, and 3) Approximately maximizing the BS's utility case, where the transmission powers of relay nodes are determined by the BS such that the BS's utility is approximately maximized. In our model, the cost incurred due to interference caused by relay nodes to uplink cellular user
communication is taken into account. The widely used Vickrey-Clarke-Groves (VCG) mechanism~\cite{RF:mascolell:microeconomic}, on which most truthful auctions designed in prior work are based~\cite{RF:bin:cao},~\cite{RF:auction:offloading},~\cite{RF:optimal:auction},~\cite{RF:HERA} (see Section~\ref{sec:related:work}),  can be applied to our network model in scenarios 1) and 2). However, in these scenarios, \emph{the VCG mechanism based auctions have several limitations, in particular, high interference cost and/ or low data rates achieved by destination nodes and low BS utility}. Also, in scenario 3, the VCG mechanism  \emph{is not applicable}. Hence, we propose novel reverse auctions for relay selection in each of the above three scenarios. Our proposed auctions for each of these three scenarios are applicable to all the above mentioned relaying schemes, viz., normal relaying, amplify-and-forward, decode-and-forward and selection relaying. We prove that all our proposed reverse auction mechanisms (i) guarantee truthful declaration of their incurred costs by relay  nodes (i.e., are \emph{incentive compatible}~\cite{RF:mascolell:microeconomic}), and (ii) satisfy the \emph{individual rationality} property~\cite{RF:mascolell:microeconomic}, i.e., the utility of a selected relay node is guaranteed to be non-negative.  Also,  we show via numerical computations that \emph{our proposed auction for the constant power case (scenario 1) outperforms the auction based on the VCG mechanism~\cite{RF:mascolell:microeconomic} in terms of achieved data rates of destination nodes, interference cost to the BS as well as BS utility; in addition, in the constant data rate case (scenario 2), the proposed auction outperforms the VCG mechanism based auction in terms of interference cost to the BS}.

The rest of this paper is organised as follows. A review of related research literature is provided in Section~\ref{sec:related:work}. Section~\ref{sec:system:model} describes our network model and game formulation and gives a brief description of various relaying schemes that a BS can employ. In Section~\ref{sec:vcg:intro}, we briefly review the VCG mechanism and explain the limitations of the auctions based on   application of the VCG mechanism to our model. In Section~\ref{sec:auctions}, we describe our proposed auctions and show that they can be truthfully implemented and satisfy individual rationality. In Section~\ref{sec:vcg}, we compare our proposed auctions with auctions based on the VCG mechanism. In Section~\ref{sec:numerical:results}, we evaluate the performance of the proposed auctions via numerical computations. We provide conclusions in Section~\ref{SC:conclusions}.

\section{Related Work}
\label{sec:related:work}
In this section, we provide a review of related research literature. Relay assisted communication is studied in~\cite{RF:Gao},~\cite{RF:Ma},~\cite{RF:Zhang}. Here, the BS encourages its users to act as relays by providing them with incentives. An auction to enable D2D sessions in cognitive mesh assisted cellular networks is proposed in~\cite{RF:Mi}. The proposed auction is proved to satisfy the individual rationality condition and can be truthfully implemented.  Auctions for data allocation in a scenario in which cellular users provide their unused data to bidders by creating Wi-Fi hotspots are studied in~\cite{RF:data:transactions}. However, the D2D communications in~\cite{RF:Gao,RF:Ma,RF:Zhang,RF:Mi,RF:data:transactions} occur overlay and thus no interference cost  is incurred to the BS. This is in contrast to the model in our paper, in which the communication between relays and destination nodes occurs underlay, and relays share channels with users that transmit on the uplink to the BS, resulting in interference cost to the BS, and hence interference management is required. 

An auction conducted by the primary user in a Cognitive Radio Network to select a relay node  to transmit its data is proposed in~\cite{RF:online:auction}. The auction is modelled as an optimal stopping problem. \ignore{where the primary user receives bid information from relays one by one and designs an optimal stopping policy.} At the stopping time, the primary user selects the relay node. It is proved that the proposed auction satisfies individual rationality and can be truthfully implemented. However, the authors did not consider the cost of interference due to deployment of the relay in the network and only considered the problem of assigning a single relay. Optimal auction based resource allocation in D2D enabled multi-tier cellular networks is studied in~\cite{RF:distributed:auction}. A higher-tier BS acts as auctioneer whereas the D2D users and lower-tier BSs bid for channels and transmission power levels. The allocation mechanism is based on allocating channels and transmission power levels such that the total data rate is maximized while minimising the interference. Our work differs from the above in that we consider both the uncertainty of information at the BS about the battery energy costs incurred by the relay nodes and the decrease in utility of the BS due to the interference caused by D2D communications. In our model, we not only consider the effect of interference by including it in the calculation of achieved data rates, but an additional loss term is introduced in the BS's utility that increases with the transmission power of a relay node. This is done to limit a relay node's transmission power.

 A double auction \ignore{based on finding a maximum matching in a bipartite graph,} for optimal assignment of relays in a cellular network consisting of multiple cellular users and multiple relay nodes is proposed in~\cite{RF:WYong}. A cellular user is assigned a relay node only when there is an increase in channel capacity by cooperation. Three assignment problems are examined: 1) Maximizing the total number of edges in a matching, 2) Maximizing the total channel capacity in the network, 3) Maximizing the social welfare in the network. A similar network setting is used in~\cite{RF:YLi} with a difference that now the energy efficiency of the source-relay-destination link is considered. A maximum matching is found to obtain an efficient relay assignment. Auctions for D2D networks are also studied in~\cite{RF:NUM}. An auction mechanism to induce truthful reporting of private local information in a D2D network scenario is proposed in~\cite{RF:NUM}. The BS  allocates resources (transmission powers) to D2D users such that the total
utility of all the D2D users is maximized. However, the auctions proposed in~\cite{RF:NUM},~\cite{RF:YLi},~\cite{RF:WYong} do not  always satisfy the incentive compatibility condition. This is in contrast to our proposed auction, which is also based on maximum matching in a bipartite graph, but is proved to satisfy the incentive compatibility condition.

Relay selection schemes in cooperative networks are studied in~\cite{RF:bin:cao},~\cite{RF:HERA}. An auction based relay assignment scheme which considers interference in the calculations of the achieved data rates in cooperative networks is  proposed in~\cite{RF:bin:cao}.  A truthful centralized single round double auction scheme to select relays is proposed  wherein the traffic flow users (source-destination pairs) and relays both submit their bids in the form of data rates achieved with and without using relays. \ignore{Based on these bids, a maximum matching algorithm to increase the total capacity of the network is used to assign the relays and the Vickrey-Clarke-Groves (VCG) mechanism is used to determine the payment to the relays.} Later a multi round auction where the relays are assigned to their buyers (cellular users) sequentially in a distributed network is proposed. However, the multi-round auction does not satisfy the incentive compatibility property.  An optimal relay assignment scheme called HERA in cooperative networks, which considers the selfish behaviour of the network users (relays) is proposed in~\cite{RF:HERA}. However, the interference caused by the relays to the BS or cellular users is not considered as the availability of orthogonal channels for relays is assumed. 
 Reverse auctions are also studied in~\cite{RF:auction:offloading} where truthfulness is achieved by following the second price auction. A relay assisted D2D communication scheme is studied in~\cite{RF:optimal:auction}, in which the BS is the auctioneer and D2D user pairs are the bidders and the BS allocates relays, channels for transmission and their respective power levels to the D2D user pairs. A D2D pair is allotted a relay if the relay results in increase in its data rate. The allocation mechanism maximizes the total increase in valuations (which depends on the achieved data rates) of all D2D user pairs. In this paper, we too consider an auction conducted by a BS to assign relay nodes to the destination nodes. However, in our work the relay nodes are selected to assist the communication between the BS and  destination nodes instead of assisting the communication between a pair of D2D users.

In the above works, the transmission power of a relay is either fixed~\cite{RF:HERA, RF:bin:cao, RF:online:auction, RF:optimal:auction, RF:WYong} or selected to satisfy a certain \emph{SINR} threshold~\cite{RF:YLi} or selected to maximize the utility function considered~\cite{RF:distributed:auction}. In contrast, in this paper,  we design truthful auctions for all of the following three scenarios: 1) the BS assigns a fixed transmission power $P$ to all the relay nodes, 2) the BS assigns transmission powers to different relay nodes to achieve the desired data rates of destination nodes, and 3) the BS selects  the transmission power of each relay node to approximately maximize the BS's utility.

Finally, the truthful auctions proposed in~\cite{RF:bin:cao},~\cite{RF:auction:offloading},~\cite{RF:optimal:auction},~\cite{RF:HERA} are based on the VCG mechanism~\cite{RF:mascolell:microeconomic}, whereas in this paper, we propose novel reverse auctions  \emph{which differ from the VCG mechanism based auctions}. Although VCG mechanism based auctions satisfy the incentive compatibility property~\cite{RF:mascolell:microeconomic}, in some contexts, they suffer from some limitations and hence alternative auctions need to be designed. In particular, the VCG mechanism selects the allocation that maximizes the social welfare~\cite{RF:mascolell:microeconomic}. In some contexts, finding this allocation is computationally infeasible.  For example, this is the case in several combinatorial auctions where multiple goods are sold simultaneously to the bidders; truthful auctions that differ from the VCG mechanism based auctions have been designed for such combinatorial auction settings in~\cite{RF:combinatorial:2},~\cite{RF:spectrum:auction}. In some contexts, the VCG mechanism based auction finds allocations with undesirable properties. For example, in keyword search auctions, where the players bid for positions in the search results, the auctioneer wants the allocation to satisfy the property that the bidders that provide the highest expected revenues occupy the top positions. But the VCG mechanism based auction does not satisfy this property in general; hence, an alternative truthful auction that satisfies this property was proposed in~\cite{RF:keyword}. However, to the best of our knowledge, \emph{our work is the first to propose truthful auctions that differ from the VCG mechanism based auctions in the context of relay selection in wireless networks}. Our proposed auctions outperform the VCG mechanism based auctions in terms of the data rates achieved by destination nodes, the utility of the BS as well as the interference cost incurred to the BS.

\section{Network Model and Problem Formulation}
\label{sec:system:model}
\subsection{Network Model}
\label{sec:network:model}
\begin{figure}[!hbt]
\centering
\includegraphics[scale=0.07]{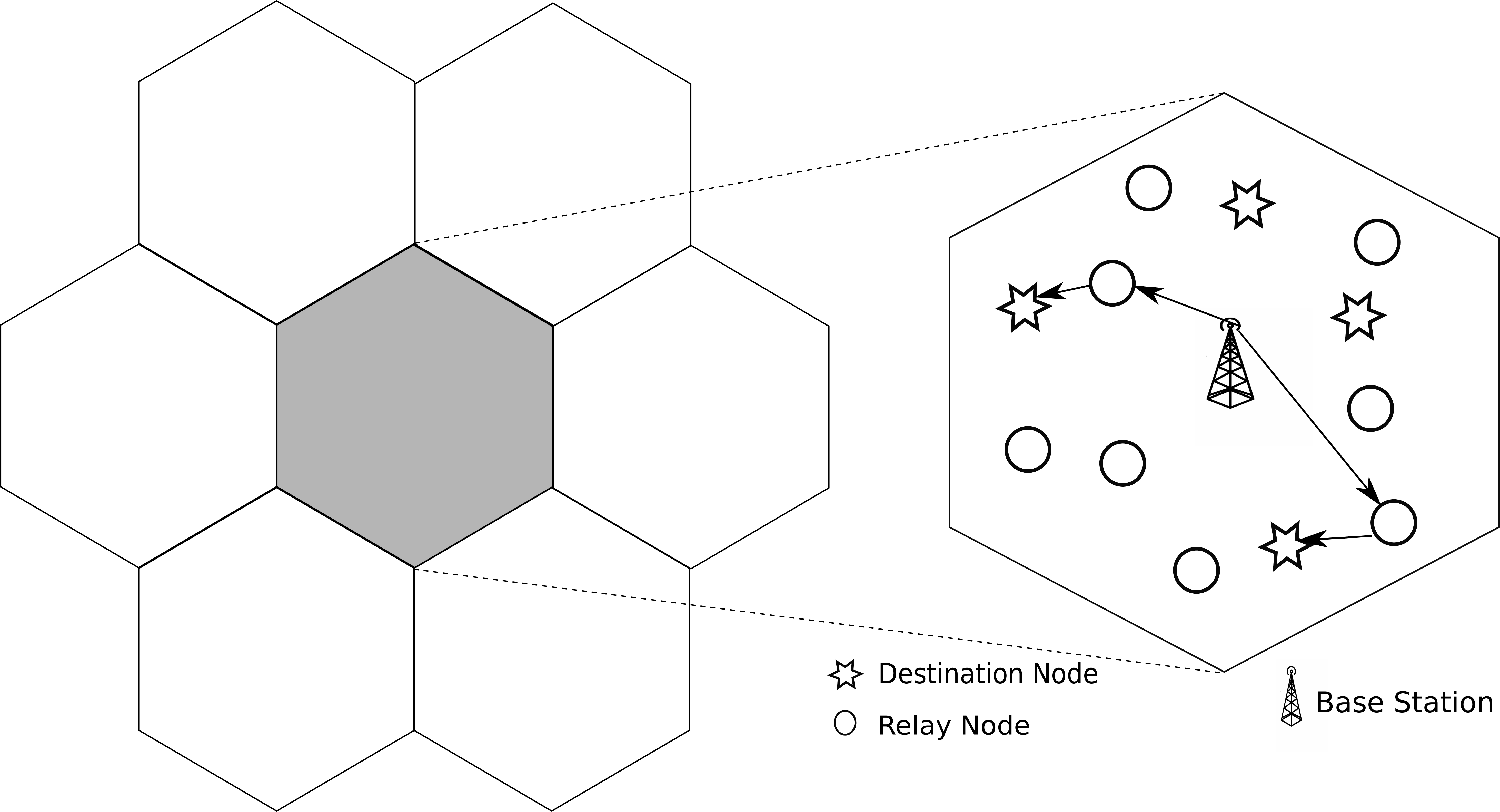}
\caption{The figure shows a cellular network with multiple cells. We assume that the BSs avoid inter-cell interference and each BS conducts an auction that assigns relay nodes present within its cell to its cellular users (destination nodes) that request relay services.}
\label{fig:network:model}
\end{figure}

We consider a cellular network with multiple cells. We assume that an interference avoidance~\cite{RF:interference} algorithm is used by the BSs, and that this algorithm assigns spectrum resources (channels) to different BSs in each time slot such that  inter-cell interference is negligible. So henceforth, we focus on a single cell which contains multiple cellular users. Fig.~\ref{fig:network:model} depicts our network model; in this figure, the cell under consideration contains cellular users shown by stars and circles. We assume that time is divided into slots and in each slot, there would be some users that would need to receive data from the BS; among them, there could be some users which request the BS for relay aided communication. Let $\mathbf{D}=\{1,\ldots,D\}$ denote the set of cellular users which request for relay services. Henceforth, we refer to the users in $\mathbf{D}$ as ``destination nodes"; these are shown by stars in Fig.~\ref{fig:network:model}. In order to deliver data to the destination nodes, the BS would send a relay request to a set of cellular users (relay nodes) in the cell which are willing to act as relays provided that they are compensated for their services. Let the set of relay nodes to which the request is sent be represented by $\mathbf{R}=\{1,2,...,R\}$; these are shown by circles in Fig.~\ref{fig:network:model}.

Information about channel conditions (qualities) is known to the BS through  Channel State Information (CSI) conveyed by the cellular users.  This CSI contains the channel gains between the BS and relays, between the BS and destination nodes and between the relays and destination nodes. This information can be estimated using reference signals, which are sent at known transmit powers are whose received powers are measured at the receivers~\cite{RF:channel:estimation}. 
For $i\in \mathbf{R}$, $j \in \mathbf{D}$, let $G_{i,j}\in K$ be the gain of the channel between relay node $i$ and destination node $j$, where $K$ represents the set of possible channel gain values, and let $G_{s,j}\in K$ be the channel gain between the BS and destination node $j$  ($s$ here represents the source which is the BS). Also, for $i\in\mathbf{R}$, let $G_{s,i}\in K$ be the gain of the channel between the BS and relay node $i$. We assume that all the above gains are known to the BS. Also, the gain, $G_{s,i}$, between the BS and relay node $i$ and for each $j \in \mathbf{D}$, the gain, $G_{i,j}$, between relay node $i$ and destination node $j$,  are known to relay node $i$. Finally, the gain, $G_{i,j}$, between each relay node $i \in \mathbf{R}$ and destination node $j$ and the gain, $G_{s,j}$, between the BS and destination node $j$  are known to destination node $j$.

Now, battery power gets consumed when a cellular user acts as a relay and it is limited. Let $B$ represent the set of all quantized battery power levels. Then, in a given time slot, a given relay node $i\in \mathbf{R}$ would be in some state $b_{i} \in B$. $B$ also includes the dead state; node $i$ cannot act as a relay if $b_i$ is the dead state. Every relay node $i\in \mathbf{R}$ knows its own battery state $b_i$. However, $b_i$ is  private information of  node $i$ and is not known to the BS.

The relays assigned to destination nodes (using an auction) reuse the channels that are used by some cellular users for \emph{uplink} (user to BS) communication and each relay node is allotted a unique channel. In particular, before the auction to assign relays to destination nodes is conducted, the BS assigns a channel  to each destination node $j\in\mathbf{D}$; this channel is also assigned to a cellular user, say $n_j$, for uplink communication. Also, a relay node which is assigned to a destination node uses this channel to communicate with its destination node. This pre-allocation of channels to destination nodes, for use by the relay nodes assigned to the destination nodes,  is useful in estimating the interference at each destination node caused by the cellular user that transmits to the BS over the uplink using the same channel. 

\subsection{Game Formulation}
\label{sec:game:formulation}

\subsubsection{Utility of a Relay Node}
Consider a relay node $i\in \mathbf{R}$ which is assigned to destination node $j\in\mathbf{D}$. Let $\Gamma_{i,j}$ be the data rate achieved at destination node $j$ with the help of relay node $i$. We assume that the payment made by the BS to the relay node is proportional to the achieved data rate $\Gamma_{i,j}$; thus, the payment made by the BS to the relay node would be $\beta \Gamma_{i,j}$, where $\beta$ is the payment per unit data rate. The utility of the relay node is given by:
\begin{equation}
\label{eq:utility1}
u_{i,j} = \beta \Gamma_{i,j} - E_{i,j},
\end{equation} 
where $E_{i,j}$ is the energy cost incurred by the relay node. The energy cost consists of two parts: 1) cost incurred while processing the received information from the BS and 2) cost incurred while transmitting the information to the destination node. Let $P_{c,i}$ denote the power required to process the received information and let $P_{i,j}$ be the power at which the relay node $i$ transmits to destination node $j$.  We assume that the total energy cost $E_{i,j}$ is a linear function of $P_{i,j}+P_{c,i}$~\footnote{All our results readily generalize to the case when $E_{i,j}=\alpha_{i} (P_{i,j}+P_{c,i}) + P_0$ where $P_0$ is a constant.} and is given by:
\begin{equation}
\label{eq:energy}
E_{i,j}=\alpha_{i} (P_{i,j}+P_{c,i}),
\end{equation}
where $\alpha_{i}$ is the cost per unit power, or, it can be said, the \emph{valuation} relay node $i \in \mathbf{R}$ has for its power. $\alpha_i$ depends on $b_i$ and is private information of node $i$. We assume that $P_{c,i}$ is proportional to the data rate, say $\Gamma_{s,i}$, of the information received by relay node $i$ from the BS, i.e., $P_{c,i}=k\Gamma_{s,i}$~\cite{RF:Christian:Isheden}; we also assume that the BS knows the constant $k$.

\subsubsection{Utility of Base Station}
Recall that we consider a cellular network in which D2D communication occurs underlay; in particular, we assume that each relay node $i$, which is assigned to a destination node, uses the same channel as some cellular user that communicates over the uplink with the BS. 

The utility of the BS is given by:
\begin{equation}
\label{eq:total:BS:utility}
U=\sum\limits_{i,j}U_{i,j},
\end{equation}
where the summation is over all relay nodes $i\in\mathbf{R}$ and destination nodes $j\in\mathbf{D}$ such that $j$ is assigned node $i$ as relay. The contribution, $U_{i,j}$, to $U$ from the pair $(i,j)$ is a function of the revenue the BS gets from $j$, the payment made to the relay node $i$ assigned to destination node $j$ and the interference caused by relay node $i$ at the BS since it uses the same channel as an uplink cellular user. Note that each destination node that receives relay service makes a payment to the BS as compensation. Let $a$ be the revenue per unit transmission rate obtained by the BS from a destination node. Also, let $C_{i,j}(P_{i,j})$ denote the cost incurred due to interference caused by relay node $i$ at the BS when relay $i$ is assigned to destination node $j$.  Then $U_{i,j}$ is given by:
\begin{equation}
\label{eq:bs:utility}
U_{i,j}=  a \Gamma_{i,j} - \beta \Gamma_{i,j} - C_{i,j}(P_{i,j}),
\end{equation}
where $\Gamma_{i,j}$ is the data rate achieved at destination node $j$ when it is assigned relay node $i$.
\begin{remark}
\label{RM:interference:cost}
As a simple example, the interference cost may be $C_{i,j}(P_{i,j})=c_iP_{i,j}$, where $c_i$ is a constant, i.e., the interference cost is a linear function of $P_{i,j}$. A more realistic expression would be $C_{i,j}(P_{i,j})=c_i(\Gamma_{n_j}-\Gamma_{n_j}^i)$, where $n_j$ denotes the cellular user on whose uplink channel relay node $i$  transmits, $\Gamma_{n_j}$ (respectively, $\Gamma_{n_j}^i$) is the data rate achieved by user $n_j$ on the uplink channel from itself to the BS when interference from relay node $i$ is absent (respectively, present). \emph{All our results, except those in Section~\ref{sec:appendix:2} where we have assumed for tractability that  $C_{i,j} (P_{i,j})=c_iP_{i,j}$, hold for arbitrary functions $C_{i,j}(P_{i,j})$.}
\end{remark}
\subsubsection{Objective}
Our objective in this paper is to design reverse auctions that can be conducted by the BS to assign to each destination node, a unique relay node. The two desirable properties of any auction are i) it must satisfy the property of individual rationality (IR)~\cite{RF:mascolell:microeconomic}, and ii) it must be truthfully implementable~\cite{RF:mascolell:microeconomic}. An auction satisfies IR if no relay gets a negative utility under any outcome of the auction~\cite{RF:mascolell:microeconomic}. Also, an auction is truthfully implementable if revealing its true valuation, $\alpha_i$, is the dominant strategy for each relay node $i \in \mathbf{R}$ \cite{RF:mascolell:microeconomic}. Our objective is to design reverse auctions, which satisfy the above two properties, for each of the following three scenarios: 
\begin{enumerate}
\item
\emph{Constant power case}, where the BS assigns a fixed transmission power $P$ to each relay node. In general, in a cellular network, the BS can either allocate different transmit power levels to different cellular users (e.g., taking into account the current channel gains) or assign a fixed transmit power to all cellular users \cite{RF:nambiar}. Although the former scheme, a variable transmit power scheme, allows a more flexible allocation, the fixed power allocation scheme is easier to implement due to its simplicity and also the loss in performance is negligible compared to the former for dense deployments of BSs~\cite{RF:j:kim},~\cite{RF:y:wei}. 
\item
\emph{Constant data rate case}, where the BS assigns a transmission power to each relay node $i$ to achieve the desired data rate, say  $\Gamma_j$, at the destination node $j$ to which $i$ is assigned. For example, when a destination node $j$ streams an audio or video file or is in an audio or video conference call, it typically requires a certain data rate $\Gamma_j$. The relay $i$ which is assigned to destination node $j$ must select its transmission power such that $j$ achieves the desired data rate $\Gamma_j$.
\item
\emph{Case where the BS selects the relay nodes's transmission powers to approximately maximize its own total utility ($U$ in \eqref{eq:total:BS:utility})}. 
\end{enumerate}
Also, for each of the above three scenarios, our objective is to design auctions 
that can be used for assignment of relays to destination nodes under each of the
following four relaying schemes-- normal relaying, amplify-and-forward, decode-and-forward and selection relaying~\cite{RF:Laneman} (see Section~\ref{subsection:relaying:schemes}).

\subsection{Relaying Schemes}
\label{subsection:relaying:schemes}
In this subsection, we briefly describe some basic cooperative communication protocols, any one of which may be employed by relay nodes assigned to destination nodes, for forwarding data. Consider a relay node $i$ which is assigned to destination node $j$. In all the following relaying schemes, we divide each time slot into two equal parts, which we denote by mini-slot 1 and mini-slot 2. In mini-slot 1, the BS transmits the message and this transmission is received by both the relay node $i$ and destination node $j$. In mini-slot 2, the relay node $i$ retransmits the message it received in mini-slot 1 (possibly after processing it), whereas the BS does not transmit any message. Depending upon the relaying scheme employed, this retransmitted signal can simply be an exact copy of the signal that relay node $i$ received in mini-slot 1 or its decoded version. The next few paragraphs give a brief overview of the operation of various relaying schemes and the data rates achieved at destination node $j$ through them.

\subsubsection{Normal Relaying Scheme}
In the normal relaying scheme, the BS transmits its message in mini-slot 1, which is received by the relay node, but ignored by the destination node; in mini-slot 2, the relay node forwards the received message to the destination. This kind of relaying operation can only extend the range of the communication or save transmission power but does not achieve any diversity gain. The data-rate capacity of this relaying scheme is determined by the weaker of the two links-- the link from the BS to the relay node and that from the relay node to the destination node. The data rate achieved at the destination node $j$ is given by:
\begin{equation}
\Gamma_{i,j} = \min \bigg\{\frac{W}{2}\log_2(1+SINR_{s,i}), \frac{W}{2}\log_2(1+SINR_{i,j})  \bigg\}.
\end{equation}
where $s$ denotes the BS, $P_s$ is the power at which the BS transmits, $W$ is the bandwidth of the channel, $SINR_{s,i}=\frac{P_{s}G_{s,i}}{I_{s,i}+N_{s,i}}$ is the signal to interference ($I_{s,i}$) plus noise ($N_{s,i}$) ratio of the link between the BS and relay node $i$, $SINR_{i,j}=\frac{P_{i,j}G_{i,j}}{I_{i,j}+N_{i,j}}$ is the signal to interference ($I_{i,j}$) plus noise ($N_{i,j}$) ratio of the link between the relay node $i$ and destination node $j$. Note that $\frac{W}{2}\log_2(1+SINR_{s,i})$ (respectively, $\frac{W}{2}\log_2(1+SINR_{i,j})$) is the Shannon capacity of the channel between the BS and relay node $i$ (respectively, relay node $i$ and destination node $j$); the factor $\frac{1}{2}$ appears in each capacity expression since communication occurs on each of the above two channels for $\frac{1}{2}$ of the duration of a time-slot.      In this work, we assume that the channel gain between the BS and the relay node is sufficiently high so that the BS can adjust its transmission power $P_{s}$ to make the capacities of both links equal. So now, the data-rate capacity equals:
\begin{equation}
 \label{eq:data:rate:normal}
\Gamma_{i,j} =  \frac{W}{2}\log_2(1+SINR_{i,j}).
\end{equation}

\vspace{1mm}
\subsubsection{Amplify-and-Forward Relaying Scheme}
The amplify-and-forward (AF) scheme  is a simple relaying scheme in which, in mini-slot $1$, the BS transmits the message to the relay and the destination node; also, the relay node amplifies the received signal and in mini-slot $2$, it forwards the amplified version of the signal to the destination node~\cite{RF:Laneman}. Apart from its simplicity and low cost, its advantage is that the relay node does not need to decode and re-encode the received signal. However, a major limitation of this scheme is that the noise in the signal received at the relay node also gets amplified. The data-rate capacity of the  AF cooperative relaying protocol is given by~\cite{RF:Laneman}:  
\begin{equation}
\Gamma_{i,j} = \frac{W}{2} \log_2 \left(1 + SINR_{s,j} + \frac{SINR_{s,i} SINR_{i,j}}{1 + SINR_{s,i} + SINR_{i,j}}\right),
\label{eq:data:rate:amplify:forward}
\end{equation}
where $SINR_{s,j}=\frac{P_sG_{s,j}}{I_{s,j}+N_{s,j}}$ is the signal to interference ($I_{s,j}$) plus noise ($N_{s,j}$) ratio of the link between the BS and destination node $j$, $SINR_{s,i}$  and $SINR_{i,j}$ are as defined above for the normal relaying scheme. 

\subsubsection{Decode-and-Forward Relaying Scheme}
In the decode-and-forward (DF) relaying scheme, in mini-slot $1$, the BS transmits the message to the relay and the destination node; the relay node decodes the received signal from the BS and re-encodes it before forwarding it to the destination node in mini-slot $2$~\cite{RF:Laneman}. As a result of decoding and encoding the received signal, the relay node incurs an additional processing cost. The data-rate capacity of this cooperative relay protocol is given by~\cite{RF:Laneman}:
\begin{align}
\Gamma_{i,j} = &\min \bigg\{\frac{W}{2}\log_2(1+SINR_{s,i}), \nonumber\\&\frac{W}{2}\log_2(1+SINR_{s,j}+SINR_{i,j})  \bigg\}, 
\label{eq:data:rate:decode:forward}
\end{align}
where the $SINR$ terms are as defined above for the AF case.

\subsubsection{Selection Relaying Scheme}
Unlike in fixed relaying schemes like AF and DF, cooperative communication is employed only if the channel conditions satisfy certain conditions in the selection relaying protocol. The BS transmits the message to the relay node and the destination node in mini-slot 1 as in the AF and DF cooperative schemes. But the relay node forwards this signal only if the SINR from the BS to the relay node is above a certain threshold $\zeta$. If this threshold constraint on the $SINR$ is satisfied, then the relay node forwards the signal using the DF protocol, otherwise the BS again transmits the same signal to the destination node in mini-slot $2$~\cite{RF:Laneman}. The data-rate capacity of the selection relaying cooperative communication protocol is given by~\cite{RF:Laneman}:
\begin{equation}
 \label{eq:data:rate:selection:relaying}
\Gamma_{i,j} =
    \begin{cases}
      \frac{W}{2}\log_2(1+2SINR_{s,j}),\hspace{-1em}& \text{if} \  SINR_{s,i} < \zeta,\\
      \frac{W}{2}\log_2(1+SINR_{s,j}+SINR_{i,j}),\hspace{-1em}& \mbox{otherwise.}
    \end{cases}
   \end{equation}
If $SINR_{s,i} < \zeta$, then relay node $i$ is not assigned to destination node $j$.

\subsection{Some  Terminology and Notations}
We now briefly explain some terminology and notations from graph theory that are used in the following sections. A graph $G=(V,E)$, with node set $V$ and edge set $E$, is a \textit{bipartite graph} if $V$ can be partitioned into two disjoint sets $V_1$ and $V_2$ such that every edge in $E$ is between a node in $V_1$ and a node in $V_2$~\cite{RF:graph:theory:west}. We represent a bipartite graph as $G=\left(V_1,V_2,E\right)$. A \textit{matching} $m\subset E$ in a bipartite graph is a collection of edges such that no two edges have a common endpoint~\cite{RF:graph:theory:west}. A matching $m$ is \textit{maximal} if $m\cup e$ is not a matching for any edge $e\in E\setminus m$~\cite{RF:graph:theory:west}.

\section{Reverse Auctions Based on the VCG Mechanism}
\label{sec:vcg:intro}
In Section~\ref{SSC:VCG:review}, we  briefly review the VCG mechanism and in Section~\ref{SSC:VCG:application}, we explain how it can be applied to our network model. In Section~\ref{SSC:VCG:limitations}, we discuss the limitations of the VCG mechanism based auctions in the context of our network model.

\subsection{Review of the VCG Mechanism}
\label{SSC:VCG:review}
The Vickrey-Clarke-Groves (VCG) mechanism~\cite{RF:mascolell:microeconomic}  is the most widely used strategy-proof method for allocation of resources and deciding on the payments to be made in standard economic models where users are rational.  Let $N$ be the set of players (agents) and $|N|=n$. Each player $i \in N$ has private information, say $\alpha_i$, called its \emph{type}. All players's types define a type vector $\alpha = (\alpha_1, \ldots, \alpha_n)$. A mechanism~\cite{RF:mascolell:microeconomic} defines a set of strategies, $A_i$, for each player $i\in N$, from which player $i$ selects a strategy $a_i$.  By the \emph{direct revelation principle}~\cite{RF:mascolell:microeconomic}, we can assume that the strategy of each player is to declare its type. Thus, the resulting strategy vector is $a=(\alpha_1,\ldots,\alpha_n )$. A mechanism computes allocation~\footnote{For example, in the context of an auction mechanism, an allocation may be a vector $Y=(Y_1,\ldots, Y_n)$, where $Y_i$ is 1 if the good is allocated to bidder $i$ and 0 else.} $o$ and payment vector $p =  (p_1,\ldots,p_n)$ as a function of strategy vector $a$. $p_i$ is the payment given to agent $i$. For each possible allocation $o$, agent $i$'s preferences are given by a valuation function $v_i(\alpha_i,o)$. If the utility of agent $i$ is denoted by $u_i(\alpha_i,a)$, an assumption required for the VCG mechanism to apply is that agents are rational and have quasi-linear utility functions of the form:
\begin{equation}
u_i(\alpha_i,a) = v_i(\alpha_i,o) + p_i.
\label{eq:ql}
\end{equation}
Under the VCG scheme, the allocation $o^*$ that satisfies the following condition is selected~\cite{RF:mascolell:microeconomic}:
\begin{equation}
\label{eq:vcg:allocation}
\sum\limits_{i=1}^{n}v_i(\alpha_i,o^*)\geq \sum\limits_{i=1}^{n}v_i(\alpha_i,o) \hspace{2mm} \forall o,
\end{equation}
and the payment $p_i$ is given by~\cite{RF:mascolell:microeconomic}:
\begin{equation}
\label{eq:vcg:payment}
p_i=\sum_{j\neq i}v_j(\alpha_j,o^*)- \sum_{j\neq i} v_j(\alpha_j,o_{-i}^*),
\end{equation}
where $o_{-i}^*$ is the allocation that would have been selected under the VCG scheme if agent $i$ did not participate in the mechanism.  

\subsection{Application of VCG Mechanism to Our Network Model}
\label{SSC:VCG:application}
The modelled game with $R$ relay nodes denoted by the set $\mathbf{R} =\{1,...,R\}$ and $D$ destination nodes denoted by the set $\mathbf{D}=\{1,\ldots,D\}$ can be described as a mechanism as follows. Each relay node $i\in \mathbf{R}$ in the network environment is an agent and has private information $\alpha_i$ (its type). In our model (see \eqref{eq:utility1} and \eqref{eq:energy}), the payment to relay node $i$ is:
\begin{equation*}
 p_i=\beta\sum\limits_{j=1}^{D} \Gamma_{i,j}y_{i,j},
 \end{equation*}
and the valuation of relay node $i$ is:
\begin{equation}
\label{eq:valuation}
v_{i}(\alpha_i,o)= -\alpha_i\sum\limits_{j=1}^{D}(P_{i,j}+P_{c,i})y_{i,j}, \text{where}
\end{equation}
\begin{equation}
y_{i,j} = 
\begin{cases}
1, & \mbox{if $i$ is assigned to $j$ under the allocation o},\\
0, & \mbox{else.}
\end{cases}
\end{equation}
Also, $\sum\limits_{j=1}^{D}y_{i,j}\leq 1$ for all $i\in\mathbf{R}$ and $\sum\limits_{i=1}^{R}y_{i,j}=1$ for all $j\in \mathbf{D}$. The inequality says that a relay may be assigned to one destination node or none and the equality says that every destination node is assigned exactly one relay. Note that the set of variables $\{y_{i,j}:i\in\mathbf{R},j\in\mathbf{D}\}$ constitute the allocation $o$. \ignore{The VCG mechanism is truthfully implementable and satisfies the property of individual rationality~\cite{RF:mascolell:microeconomic}. }

We now apply the VCG mechanism to the following two scenarios: (A) Constant power case and (B) Constant data rate case.

(A) Constant power case:
In this scenario, by \eqref{eq:vcg:allocation} and \eqref{eq:valuation}, under the VCG mechanism, relay nodes are assigned to destination nodes such that the following expression is minimized:
\begin{equation}
\sum\limits_{j=1}^D\sum\limits_{i=1}^R\alpha_i(P+P_{c,i})y_{i,j},
\label{eq:vcg:constant:power}
\end{equation}
where $\sum\limits_{j=1}^{D}y_{i,j}\leq 1$ for all $i\in\mathbf{R}$ and $\sum\limits_{i=1}^{R}y_{i,j}=1$ for all $j\in \mathbf{D}$. Also, under the VCG mechanism, the payment to every selected relay node is (from \eqref{eq:vcg:payment}) $\alpha_l (P+P_{c,l})$, where $\alpha_l(P+P_{c,l})$ is the $(D+1)$'st lowest value from the set $\{\alpha_m(P+P_{c,m}):m\in\mathbf{R}\}$. The rest of the nodes get a payment of $0$. Note that every relay node that is selected under the VCG mechanism is paid the same amount $\alpha_l(P+P_{c,l})$. 

\begin{theorem}
\label{TH:VCG:const:power}
The above VCG mechanism based auction satisfies individual rationality and can be truthfully implemented.
\end{theorem}

The claim in Theorem~\ref{TH:VCG:const:power} that the VCG mechanism based auction can be truthfully implemented  follows from  Proposition 23.C.4 in~\cite{RF:mascolell:microeconomic}; also, since by \eqref{eq:vcg:payment}, every relay node gets a non-negative utility, the above auction also satisfies individual rationality.

We now evaluate the time complexity of the  above auction.

\begin{proposition}
\label{prop:time:complexity:vcg:const:power}
The time complexity of the above VCG mechanism based auction is $O(R\log R)$.
\end{proposition}

\begin{IEEEproof}
Relay nodes are assigned to destination nodes such that the expression in \eqref{eq:vcg:constant:power} is minimized. This is equivalent to selecting the $D$ relay nodes with the smallest values of $\alpha_{i}(P+P_{c,i})$ and assigning each of these selected relay nodes to any one destination node. This can be achieved by sorting the list of the $\alpha_i(P+P_{c,i})$ values of all the relays $i$ in ascending order and selecting the first $D$ relays from this list. The complexity of sorting a list of $R$ values (e.g., using the Merge sort algorithm~\cite{RF:algorithms}) is $O(R\log R)$. Next, the payment to each selected relay is $\alpha_m(P+P_{c,m})$, where $m$ is the relay corresponding to the $(D+1)$'st value in the above sorted list; this payment can be found in constant time. The result follows. 
\end{IEEEproof}

(B) Constant data rate case:
In this case, \ignore{the proposed auction assigns relay nodes to destination nodes such that the following expression is minimized (see Fig.~\ref{algorithm:auction:constant:data:rate}):
\begin{equation}
\sum\limits_{j=1}^D\sum\limits_{i=1}^R\frac{\alpha_i(P_{i,j}+P_{c,i})}{\Gamma_{j}}y_{i,j}. 
\label{eq:proposed:constant:data:rate}
\end{equation}
But} by \eqref{eq:vcg:allocation} and \eqref{eq:valuation}, under the VCG mechanism, relay nodes are assigned to destination nodes such that the following expression is minimized:
\begin{equation}
\sum\limits_{j=1}^D\sum\limits_{i=1}^R\alpha_i(P_{i,j}+P_{c,i})y_{i,j},
\label{eq:vcg:constant:data:rate}
\end{equation}
where $\sum\limits_{j=1}^{D}y_{i,j}\leq 1$ for all $i\in\mathbf{R}$ and $\sum\limits_{i=1}^{R}y_{i,j}=1$ for all $j\in \mathbf{D}$. \ignore{It can be easily verified that the assignment of relay nodes to destination nodes under the proposed auction may differ from that under the VCG mechanism.} Also, the payment to each relay node is calculated using \eqref{eq:vcg:payment}.

\begin{theorem}
\label{TH:VCG:auction:const:data:rate}
The above VCG mechanism based auction satisfies individual rationality and can be truthfully implemented.
\end{theorem}

The claim in Theorem~\ref{TH:VCG:auction:const:data:rate} that the VCG mechanism based auction can be truthfully implemented  follows from  Proposition 23.C.4 in~\cite{RF:mascolell:microeconomic}; also, since by \eqref{eq:vcg:payment}, every relay node gets a non-negative utility, the above auction also satisfies individual rationality.

We now evaluate the time complexity of the above auction. We write the computational complexity of this auction in terms of the computational complexity of the \emph{Hungarian algorithm}~\cite{RF:Hungarian}, which can be used to find the minimum weighted maximal matching in a bipartite graph $G=(\mathbf{R},\mathbf{D},E)$. The Hungarian algorithm has a time complexity of $O((R+D)^2\log(R+D)+(R+D)RD)$~\cite{RF:combinatorial:optimization}, where $R=\left|\mathbf{R}\right|$ and $D=\left|\mathbf{D}\right|$. Let $\mathcal{H}$ denote this time complexity.

\begin{proposition}
\label{prop:time:comp:vcg:const:data}
The time complexity of the above VCG mechanism based auction is $O(D\mathcal{H})$.
\end{proposition}

\begin{IEEEproof}
Relay nodes are assigned to destination nodes such that the expression in \eqref{eq:vcg:constant:data:rate} is minimized. To find this assignment, we construct a complete bipartite graph $G=(\mathbf{R},\mathbf{D},E)$, where the edge weight between relay node $i$ and destination node $j$ is $\alpha_i(P_{i,j}+P_{c,i})$. Relays are assigned to destination nodes by finding the maximal matching with the minimum weight. This is done using the Hungarian algorithm~\cite{RF:Hungarian}, which has a time complexity of $O(\mathcal{H})$.

Next, we find the time complexity of computing the payment $p_i$ to relay node $i$ (see \eqref{eq:vcg:payment}). To find the allocation $o_{-i}^*$ in \eqref{eq:vcg:payment}, we remove node $i$ and all its incident edges from the graph $G$ and find the maximal matching with the minimum weight in the resultant graph using the Hungarian algorithm; the time complexity of this computation is $O(\mathcal{H})$. The allocation $o_{-i}^*$ needs to be found for every relay node $i$ that is assigned to a destination node. Since $D$ relay nodes are assigned to destination nodes, the complexity of the VCG mechanism based auction is  $O(D\mathcal{H})$. 

The result follows.
\end{IEEEproof}

\subsection{Limitations of  VCG Mechanism Based Auctions in the Context of our Network Model}
\label{SSC:VCG:limitations}
\subsubsection{Constant Power Case}
From \eqref{eq:vcg:constant:power}, it can be seen that in the constant power case,  the VCG mechanism based auction selects the $D$ nodes in $\mathbf{R}$ with the $D$ smallest values of the quantity $\alpha_i(P+P_{c,i})$ as relays. The outcome of the VCG mechanism based auction may be any arbitrary assignment of the $D$ relay nodes in $\mathbf{R}$ with the $D$ smallest values of $\alpha_i(P+P_{c,i})$ to the nodes in $\mathbf{D}$; note that every such assignment minimizes the quantity in \eqref{eq:vcg:constant:power}. Thus, the \emph{VCG mechanism based auction ignores the data rates achieved by the destination nodes; hence, the achieved data rates of the destination nodes under the VCG mechanism based auction are lower than those under our proposed auction} (which will be described in Section~\ref{subsec:constant:power} and which takes the data rates achieved by destination nodes into account while assigning relays to destination nodes).  Also, by \eqref{eq:total:BS:utility} and \eqref{eq:bs:utility}, the utility
of the BS is an increasing function of the achieved data rates
of the destination nodes; hence, \emph{the utility of the BS under the VCG mechanism based auction is lower than that under our proposed auction.} Another limitation of the VCG mechanism based auction is that it \emph{completely ignores the interference cost to the BS; this results in a higher interference cost to the BS under the VCG mechanism based auction than under our proposed auction.} The above limitations of the VCG mechanism based auction are illustrated by the following simple example.

\begin{example}
\label{ex:const:power}
Suppose there are three potential relays, say $\{1,2,3\}$, and one destination node, say $1$. Suppose the data processing costs are zero, i.e.,  $P_{c,i}=0$ for every relay $i$ (see \eqref{eq:energy}); also, $a=2$ (see \eqref{eq:bs:utility}). Let $\alpha_{1}=1$, $\alpha_{2}=1.1$ and $\alpha_{3}=2$. Suppose each  relay  transmits at the constant power level of $P=0.25 W$ and let the corresponding data rates achieved at the destination node be $\Gamma_{1,1}=1$ Mbps and $\Gamma_{2,1}=\Gamma_{3,1}=5$ Mbps. Let the interference costs to the BS (see \eqref{eq:bs:utility}) be $C_{1,1}(P)=1$, $C_{2,1}(P)=0.5$ and $C_{3,1}(P)=3$. 

By \eqref{eq:vcg:constant:power}, the VCG mechanism assigns relay $1$ to the destination node and the corresponding data rate achieved by the destination node is  $\Gamma_{1,1}=1$ Mbps. However, it can be checked that under our proposed auction, relay $2$ is assigned to the destination node and  the data rate achieved by the destination node is $\Gamma_{2,1}=5$ Mbps. Also, the utility of the BS under the VCG mechanism based auction (respectively,  proposed auction) is  $0.725$ (respectively, $8.25$) (see \eqref{eq:bs:utility}, \eqref{eq:vcg:payment}). Finally, the interference cost to the BS under the VCG mechanism based auction (respectively,  proposed auction)  is $C_{1,1}(P)=1$ (respectively, $C_{2,1}(P)=0.5$). Thus, the proposed auction significantly outperforms the VCG mechanism based auction in terms of the data rate achieved by the destination node, BS utility as well as interference cost incurred to the BS.
\end{example}

\subsubsection{Constant Data Rate Case}
In this case, a limitation of the VCG mechanism based auction is that it \emph{completely ignores the interference cost to the BS; this results in a higher interference cost to the BS under the VCG mechanism based auction than under our proposed auction (see Section~\ref{subsec:constant:data:rate})}. The above limitation of the VCG mechanism based auction is illustrated by the following simple example. 

\begin{example}
\label{ex:const:data}
Suppose there is one destination node, say $1$, and three potential relay nodes $\{1,2,3\}$. The destination node requests a data rate of 3 Mbps. Suppose the data processing costs are zero, i.e.,  $P_{c,i}=0$ for every relay $i$ (see \eqref{eq:energy}); also, $a=2$ (see \eqref{eq:bs:utility}). Let the power required by  relays $1$, $2$ and $3$ to achieve the requested data rate be $P_{1,1} = 0.5$, $P_{2,1} = 0.7$ and $P_{3,1} = 1$ W respectively. Let $\alpha_{1}=1$, $\alpha_{2}=1.1$ and $\alpha_{3}=2.5$. Let the interference costs incurred to the BS be $C_{1,1}(0.5)=3$, $C_{2,1}(0.7)=C_{3,1}(1)=1$. In this case, the VCG mechanism based auction assigns relay $1$ to the destination node (see \eqref{eq:vcg:constant:data:rate}), whereas it can be checked that our proposed auction assigns relay $2$ to the destination node. The resultant  interference cost to the BS under the VCG mechanism based auction (respectively, proposed auction) is $C_{1,1}(0.5)=3$ (respectively, $C_{2,1}(0.7)=1$). Thus, the proposed auction significantly outperforms the VCG mechanism based auction in terms of the interference cost incurred to the BS.
\end{example}

\subsubsection{Selecting Power to Approximately Maximize the Utility of the BS}
In this case, \emph{the VCG mechanism is not applicable since it does not specify how the transmission powers of the relays should be set so as to approximately maximize the BS's utility.}

\section{Proposed Reverse Auctions}
\label{sec:auctions}
In Sections~\ref{subsec:constant:power},~\ref{subsec:constant:data:rate} and~\ref{subsec:maximize:BS:utility}, we present our proposed auctions for the constant power case, constant data rate case and the case where the transmit powers are selected to approximately maximize the BS's utility respectively. In Section~\ref{sec:appendix:2}, we provide expressions for the transmission power of a relay node under various relaying schemes for the constant data rate scenario and approximate BS utility maximization scenario.

Recall that the BS assigns every destination node $j\in\mathbf{D}$ a channel on which the relay assigned to $j$ transmits; this channel is also assigned to a cellular user, say $n_j$, for uplink communication. If destination node $j$ is assigned relay node $i$, then we let $\Gamma_{n_j}$ (respectively, $\Gamma_{n_j}^i$) denote the data rate achieved by cellular user $n_j$ on the uplink channel from itself to the BS when interference from relay node $i$ is absent (respectively, present).

\subsection{Constant Power Case}
\label{subsec:constant:power}
In this subsection, we consider the case where the BS assigns a fixed transmission power $P$ to all the relay nodes. By \eqref{eq:utility1} and \eqref{eq:energy}, the utility of a relay node $i$ if it is assigned to destination node $j$ is:
\begin{equation}
u_{i,j}=\beta_i\Gamma_{i,j}-\alpha_i(P+P_{c,i}).
\label{eq:utility:constant:power}
\end{equation}
A relay node gets $0$ utility if it is not assigned to any destination node. We now propose an auction which is based on matching in bipartite graphs. First, each relay $i$ declares its valuation, $\alpha_i$, to the BS. Then we construct a complete bipartite graph~\footnote{A bipartite graph $G=(V_1,V_2,E)$ is said to be complete if there is an edge between every $v_1\in V_1$ and every $v_2\in V_2$.} $G=(\mathbf{R},\mathbf{D},E)$, where $\mathbf{R}$ (respectively, $\mathbf{D}$) is the set of all relay nodes (respectively, destination nodes). The weight of the edge between relay node $i \in \mathbf{R}$ and destination node $j \in \mathbf{D}$ is defined to be $\frac{\alpha_i(P+P_{c,i})}{\Gamma_{i,j}}$ if $C_{i,j}(P)\leq C_T$, else $\infty$, where $C_{i,j}(\cdot)$ is as in \eqref{eq:bs:utility} and $C_T$ is a parameter.   Let $(i,j)$ denote the edge between relay node $i\in \mathbf{R}$ and destination node $j\in \mathbf{D}$. Also, let $\mathbf{M}$ denote the set of all possible maximal matchings in the above graph. For every maximal matching $m\in\mathbf{M}$, we define a corresponding weight $w_m$, which is equal to the sum of weights of all the edges in $m$. Let $R_m$ denote the set of all relay nodes which are in the neighbourhood~\footnote{The neighbourhood of a vertex $v$ under the matching $m$ is the set of all vertices which are connected by an edge in $m$ with the vertex $v$.  The neighbourhood of a set of vertices $C$ under the matching $m$ is the set $\{v:\mbox{$v$ is in the neighbourhood of a vertex $c\in C$ under $m$}\}$.} of $\mathbf{D}$ under the matching $m$. The proposed algorithm is based on finding a maximal matching with the minimum weight. If we denote $w_{min}=\min\limits_{m\in \mathbf{M}}w_m$ and $m_{min}=\argmin\limits_{m\in \mathbf{M}}w_m$, we select the relay nodes $R_{m_{min}}$ as the auction winners, each of which is assigned to its neighbour in the set $\mathbf{D}$ under the matching $m_{min}$. 

We denote for every relay node $i$, $w_{m_{min}^{-i}}=\min\limits_{m\in \mathbf{M}, i\notin R_m}w_m$ and $\mathbf{M}_i$ as the set of all maximal matchings such that for every $m\in \mathbf{M}_i$, we have $i\in R_m$ and $w_m\leq w_{m_{min}^{-i}}$. If a relay node $i\in R_{m_{min}}$, then for every $m\in \mathbf{M}_i$, we define:
\begin{equation}
p_{i,m}=\left(w_{m_{min}^{-i}}-w_m+\frac{\alpha_i(P+P_{c,i})}{\Gamma_{i,j}}\right)\Gamma_{i,j},
\label{eq:pim}
\end{equation} 
where $j\in \mathbf{D}$ is the adjacent vertex of node $i$ in matching $m$. The payment given to relay node $i$ is $p_i=\max\limits_{m\in \mathbf{M}_i} p_{i,m}$. 

The sequence of steps that implements the above auction is provided in Fig.~\ref{algorithm:auction:constant:power}. Note that the weight of the edge between $i$ and $j$ is defined to be $\infty$ if $C_{i,j}(P) > C_T$ (see step 2 of Fig.~\ref{algorithm:auction:constant:power}) so as to ensure that only those allocations of relays to destination nodes for which the cost incurred due to
interference caused by relays at the BS is sufficiently low can possibly be selected.

\begin{figure}[!hbt]
\mbox{}\hrulefill \\
\begin{scriptsize}
\begin{algorithmic}[1]
\STATE{Construct a complete weighted bipartite graph $G=(\mathbf{R},\mathbf{D},E)$.}
\STATE{Define the weight of edge $(i,j)$ to be $\frac{\alpha_i(P+P_{c,i})}{\Gamma_{i,j}}$ if $C_{i,j}(P)\leq C_T$, else $\infty$.}
\STATE{Select a maximal matching $m_{min}$ such that $m_{min}=\argmin\limits_{m\in \mathbf{M}}w_m$.}
\STATE{If $(i,j)\in m_{min}$, where $(i,j)\in\mathbf{R}\times\mathbf{D}$, assign relay node $i$ to destination node $j$.}
\STATE{If relay node $i$ is assigned to destination node $j$, then it is paid $p_i=\max\limits_{m\in \mathbf{M}_i} p_{i,m}$, else $p_i=0$ and relay node $i$ is not required to transmit any data.}
\end{algorithmic}
\end{scriptsize}
\mbox{}\hrulefill
\caption{Auction for constant power case.}
\label{algorithm:auction:constant:power} 
\end{figure}

\begin{theorem}
\label{proposition:constant:power}
The auction in Fig.~\ref{algorithm:auction:constant:power} satisfies individual rationality and can be truthfully implemented.
\end{theorem}
\begin{IEEEproof}
Let us consider relay node $i\in \mathbf{R}$. We denote $w_{min}^i=\min\limits_{m\in \mathbf{M}, i\in R_m}w_m$.  Let us assume that node $i$ is not selected as a relay when it reveals its valuation $\alpha_i$ truthfully. This implies that $w_{min}^i\geq w_{min}$. Assume that instead it declares $\alpha_i'$. This leads to a change in the values of $w_m$, $m\in \mathbf{M}$. As a result, let $w_m'$ denote the new weight of the maximal matching $m\in \mathbf{M}$. If $\alpha_i' > \alpha_i$, then $w_{min}=w_{min}'\leq w_{min}^i < w_{{min}}'^{i}$. So node $i$ is still not selected as a relay. If $\alpha_i'\leq \alpha_i$, then node $i$ is selected if $w_{{min}}'^{i}\leq w_{min}$. Let $M_i'=\{m\in \mathbf{M}:i\in R_m, w'_m\leq w_{m_{min}^{-i}}' \}$ (Note that in this case, $w_{m_{min}^{-i}}'=w_{min}$). The payment to node $i$ is $p_i'=\max\limits_{m\in \mathbf{M}_i'} p_{i,m}'$. But for every $m\in \mathbf{M}_i'$, we have:
\begin{align*}
p_{i,m}'&=\left(w'_{m_{min}^{-i}}-w'_{m}+\frac{\alpha_i'(P+P_{c,i})}{\Gamma_{i,j}}\right)\Gamma_{i,j}\\
&=\left(w_{min}-w'_{m}+\frac{\alpha_i'(P+P_{c,i})}{\Gamma_{i,j}}\right)\Gamma_{i,j},
\end{align*}
where $(i,j)\in m$. But we have $w_m-\frac{\alpha_i(P+P_{c,i})}{\Gamma_{i,j}}=w'_{m}-\frac{\alpha_i'(P+P_{c,i})}{\Gamma_{i,j}}$. Substituting this in the above equality, we get
\begin{align*}
p'_{i,m}&=(w_{min}-w_m)\Gamma_{i,j}+\alpha_i(P+P_{c,i})\\
&\leq\alpha_i(P+P_{c,i}).
\end{align*}
The above inequality holds because $w_{min}\leq w_m$. Since $p_i'=\max\limits_{m\in \mathbf{M}_{i}'}p_{i,m}$ and by \eqref{eq:utility:constant:power}, it follows that the utility of node $i$ is $\leq 0$ when it falsely declares its valuation to be $\alpha_i'$. Now, let us consider the case where relay node $i$ is selected and is assigned to destination node $j$ when it declares its valuation $\alpha_i$ truthfully. Suppose node $i$ declares $\alpha_i'$ instead and is still selected as a relay. Then $p'_{i,m}$ for each $m$ is equal to $(w_{m_{min}^{-i}}-w_m'+\frac{\alpha_i'(P+P_{c,i})}{\Gamma_{i,j}})\Gamma_{i,j}$. But as stated above, we have $w_m-\frac{\alpha_i(P+P_{c,i})}{\Gamma_{i,j}}=w_{m}'-\frac{\alpha_i'(P+P_{c,i})}{\Gamma_{i,j}}$. So $p_{i,m}'=p_{i,m}$. By separately considering the cases $\alpha_i^{\prime} < \alpha_i$ and $\alpha_i^{\prime} > \alpha_i$, it can be checked that this implies that a node $i$ which is selected as a relay when it reveals its true valuation will not get any additional benefit by manipulating its valuation. Also, if node $i$ is selected as a relay when it reveals its true valuation, then the payment made to it is $p_i=\max\limits_{m\in \mathbf{M}_i} p_{i,m}$. Since $p_{i,m}=\left(w_{m_{min}^{-i}}-w_m+\frac{\alpha_i(P+P_{c,i})}{\Gamma_{i,j}}\right)\Gamma_{i,j}$ and $w_m\leq w_{m_{min}^{-i}}$ for $m\in \mathbf{M}_i$, we have $p_{i,m}\geq\alpha_i(P+P_{c,i})$. So by \eqref{eq:utility:constant:power}, the utility of node $i$ is $\geq 0$. This proves the individual rationality property. The result follows.
\end{IEEEproof}

\begin{remark}
Note that in the above auction, an expression for the data rate $\Gamma_{i,j}$ is not mentioned. The BS can choose the type of relaying scheme it wants to implement and the data rate expression is chosen accordingly. For example, if the BS chooses the decode-and-forward relaying scheme, then the data rate expression in \eqref{eq:data:rate:decode:forward} is used to calculate the achieved data rate at the destination node for each of the relay nodes in $\mathbf{R}$. It can be checked that Theorem~\ref{proposition:constant:power} and its proof hold regardless of which of the four relaying schemes described in Section~\ref{subsection:relaying:schemes} is used.
\end{remark}


\begin{proposition}
\label{prop:time:constant:power}
The  time complexity of the auction in Fig.~\ref{algorithm:auction:constant:power} is  $O(D^2\mathcal{H})$.
\end{proposition}

\begin{IEEEproof}
Relays are assigned to destination nodes by finding the maximal matching of the bipartite graph $G=(\mathbf{R},\mathbf{D},E)$ with minimum weight (see step 3 in Fig.~\ref{algorithm:auction:constant:power}). This can be done using the Hungarian algorithm~\cite{RF:Hungarian}, which has a time complexity of $O(\mathcal{H})$.

Next, we find the time complexity of computing the payment $p_i$ made to a relay node $i$ assigned to destination node $j$ (see step 5 in Fig.~\ref{algorithm:auction:constant:power}). Let $\mathbf{M}_i^{k}$ denote the set of all maximal matchings $m$ such that $m$ contains the edge $(i,k)$ and $w_m\leq w_{m_{min}^{-i}}$. Then we have $\mathbf{M}_i=\cup_{k\in\mathbf{D}}\mathbf{M}_i^{k}$. Note that $p_i=\max\limits_{m\in \mathbf{M}_i} p_{i,m} = \max\limits_{k\in \mathbf{D}}\left(\max\limits_{m\in \mathbf{M}_i^{k}}p_{i,m}\right)$. Let $p_i^k=\max\limits_{m\in \mathbf{M}_i^{k}}p_{i,m}$. Then
by \eqref{eq:pim}, we can write: 
\begin{equation}
p_i^k=w_{m_{min}^{-i}}\Gamma_{i,k}+\alpha_i(P+P_{c,i})-\Gamma_{i,k}\min\limits_{m\in \mathbf{M}_i^k}w_m.
\label{eq:pi}
\end{equation}
Next, $\min\limits_{m\in \mathbf{M}_i^k}w_m$ for a given destination node $k\in\mathbf{D}$  can be found as follows. Find the maximal matching of the complete bipartite graph $G^{-\{i,k\}}=\left(V\setminus\{i,k\},E^{-(i,k)}\right)$, where $E^{-(i,k)}=\{e\in E:e\neq (i,l),(l,k)\forall l\in\mathbf{R}\cup\mathbf{D}\}$, with minimum weight. This can done using the Hungarian algorithm on the graph $G^{-\{i,k\}}$.
Let $m_{min}^{G^{-\{i,k\}}}$ denote the maximal matching of $G^{-\{i,k\}}$ with the least weight and let $w_{min}^{G^{-\{i,k\}}}$ denote the weight of this matching. If $w_{min}^{G^{-\{i,k\}}}\leq w_{m_{min}^{-i}}-\frac{\alpha_iP}{\Gamma_{i,k}}$, then let $\min\limits_{m\in \mathbf{M}_i^k}w_m = w_{min}^{G^{-\{i,k\}}} + \frac{\alpha_iP}{\Gamma_{i,k}}$, else let $\min\limits_{m\in \mathbf{M}_i^k}w_m = \infty$.   Next, substituting the value of $\min\limits_{m\in \mathbf{M}_i^k}w_m$ into \eqref{eq:pi}, $p_i^k$ can be found. Finally, we calculate the payment as $p_i = \max\limits_{k\in \mathbf{D}} p_i^k$. Since for calculating the payment $p_i$ to relay node $i$, we run the Hungarian algorithm on the bipartite graph $G^{-\{i,k\}}$ for every $k \in \mathbf{D}$, the time complexity of computing the payment $p_i$ is $O(D\mathcal{H})$. Since the payment needs to be computed for each of the $D$ relay nodes that are assigned to destination nodes, the overall time complexity is $O\left(D^2\mathcal{H}\right)$. The result follows.
\end{IEEEproof}

\subsection{Constant Data Rate Case}
\label{subsec:constant:data:rate}
The auction for this case is similar to the auction that is proposed for the constant power case, with the difference being that instead of assigning a constant power $P$ for each of the relays, the BS now assigns a power $P_{i,j}$ to relay $i$ assigned to destination node $j$ such that $\Gamma_{i,j}=\Gamma_{j}$. If the required power $P_{i,j}>P_m$, where $P_m$ is the maximum transmission power of a relay node, then we do not assign relay node $i$ to destination node $j$. Closed form expressions for the power $P_{i,j}$ for each of the four relaying schemes described in Section~\ref{subsection:relaying:schemes} in the case when the function $C_{i,j}(P_{i,j})$ in  \eqref{eq:bs:utility} equals $c_iP_{i,j}$ are provided in Section~\ref{sec:appendix:2}. Similar to the constant power case, after each relay declares its valuation, $\alpha_{i}$, to the BS, we construct a complete bipartite graph $(\mathbf{R},\mathbf{D},E)$; the weight of the edge between relay node $i$ and destination node $j$ is $\frac{\alpha_i(P_{i,j}+P_{c,i})}{\Gamma_j}$ if $P_{i,j}\leq P_m$ and $C_{i,j}(P_{i,j})\leq C_{T}$ else $\infty$ \ignore{if $P_{i,j}>P_m$}. The payment to relay node $i$ if it is assigned to destination node $j$ is given by $p_i=\max\limits_{m\in \mathbf{M}_i}p_{i,m}$, where:
\begin{equation}
p_{i,m}=\left(w_{m_{min}^{-i}}-w_m+\frac{\alpha_i(P_{i,j}+P_{c,i})}{\Gamma_{j}}\right)\Gamma_{j}
\label{eq:pim:const:data}
\end{equation} 
The sequence of steps that implements the proposed auction is given in Fig.~\ref{algorithm:auction:constant:data:rate}.

\begin{figure}[!hbt]
\mbox{}\hrulefill \\
\begin{scriptsize}
\begin{algorithmic}[1]
\STATE{Construct a complete weighted bipartite graph $G=(\mathbf{R},\mathbf{D},E)$.}
\STATE{Define the weight of edge $(i,j)$ to be $\frac{\alpha_i(P_{i,j}+P_{c,i})}{\Gamma_{j}}$ if $C_{i,j}(P_{i,j})\leq C_{T}$ and $P_{i,j}\leq P_m$, else $\infty$.}
\STATE{Select a maximal matching $m_{min}$ such that $m_{min}=\argmin\limits_{m\in \mathbf{M}}w_m$.}
\STATE{If $(i,j)\in m_{min}$, where $(i,j)\in\mathbf{R}\times\mathbf{D}$, assign relay node $i$ to destination node $j$.}
\STATE{If relay node $i$ is assigned to destination node $j$, then it is paid $p_i=\max\limits_{m\in \mathbf{M}_i} p_{i,m}$, else $p_i=0$ and relay node $i$ is not required to transmit any data.}
\end{algorithmic}
\end{scriptsize}
\mbox{}\hrulefill
\caption{Auction for constant data rate case.}
\label{algorithm:auction:constant:data:rate} 
\end{figure}

\begin{theorem}
\label{proposition:constant:data:rate}
The auction in Fig.~\ref{algorithm:auction:constant:data:rate} satisfies individual rationality and can be truthfully implemented.
\end{theorem}

The proof is similar to that of Theorem~\ref{proposition:constant:power} and is omitted for brevity. Also, it can be checked that Theorem~\ref{proposition:constant:data:rate} holds regardless of which of the four relaying schemes described in Section~\ref{subsection:relaying:schemes} is used. 
\begin{proposition}
The  time complexity of the auction in Fig.~\ref{algorithm:auction:constant:data:rate} is  $O(D^2\mathcal{H})$.
\end{proposition}

The proof is similar to that of Proposition~\ref{prop:time:constant:power} and is omitted for brevity.

\subsection{Selection of Power to Approximately Maximize BS Utility}
\label{subsec:maximize:BS:utility}
In this subsection, we design an auction for the case in which the BS requests each relay node to transmit at a power that will approximately maximize the BS's utility. Let $P_{i,j}$ denote the power at which the BS requires relay node $i$ to transmit to destination node $j$~\footnote{$P_{i,j}=0$ if relay node $i$ is not assigned to destination node $j$.}. The BS makes a payment of $\beta_{i,j}\Gamma_{i,j}$ if relay node $i$ is assigned to destination node $j$. So by \eqref{eq:utility1} and \eqref{eq:energy} the utility of relay node $i$ is given by:
\begin{equation}
\label{eq:max:relay:utility:1}
u_{i,j}=\beta_{i,j}\Gamma_{i,j}-\alpha_i(P_{i,j}+P_{c,i}),
\end{equation}
and by \eqref{eq:bs:utility} the contribution to the utility of the BS ($U$ in \eqref{eq:total:BS:utility}) from pair $(i,j)$ when relay node $i$ is assigned to destination node $j$ is given by:
\begin{equation}
\label{eq:BS:utility1}
U_{i,j}=(a-\beta_{i,j})\Gamma_{i,j}-C_{i,j}(P_{i,j}).
\end{equation}
For the individual rationality condition to be satisfied, $u_{i,j}\geq 0$ $\forall i,j$; so by \eqref{eq:max:relay:utility:1}, $\beta_{i,j}\Gamma_{i,j}\geq \alpha_i(P_{i,j}+P_{c,i})$. However, by \eqref{eq:BS:utility1} the BS gets maximum utility when $\beta_{i,j}\Gamma_{i,j}=\alpha_i(P_{i,j}+P_{c,i})$. So the maximum contribution to the utility of the BS from pair $(i,j)$ when relay $i$ is assigned to destination node $j$ and relay $i$ transmits at power $P_{i,j}$ is:
\begin{equation}
\label{eq:max:BS:utility:2}
U_{i,j}=a\Gamma_{i,j}-\alpha_i(P_{i,j}+P_{c,i})-C_{i,j}(P_{i,j}).
\end{equation}
Since the only variable in the above expression is $P_{i,j}$, we find the power that maximizes $U_{i,j}$. Suppose $P_{i,j}^*$\ignore{~\footnote{In case $P_{i,j}^*\leq0$, then we choose $P_{i,j}=0$, else $P_{i,j}=P_{i,j}^*$.}} maximizes $U_{i,j}$ in \eqref{eq:max:BS:utility:2}, i.e.,
\begin{equation}
P_{i,j}^*=\argmax_{0\leq P_{i,j}\leq P_m} \left(a\Gamma_{i,j}-\alpha_i(P_{i,j}+P_{c,i})-C_{i,j}(P_{i,j})\right).
\label{eq:U:maximizer}
\end{equation} 
$P_{i,j}^*$ depends on the type of relaying scheme employed by the BS. Closed form expressions for the power $P_{i,j}^*$ for each of the four relaying schemes described in Section~\ref{subsection:relaying:schemes} are provided in Section~\ref{sec:appendix:2} for the case when $C_{i,j}(P_{i,j})=c_iP_{i,j}$. The maximum contribution to the utility of the BS from pair $(i,j)$ when relay node $i$ is assigned to destination node $j$ is:
\begin{equation}
U_{i,j}^*=a\Gamma_{i,j}^*-\alpha_i(P_{i,j}^*+P_{c,i})-C_{i,j}(P_{i,j}^*)
\label{eq:BS:utility:final}
\end{equation}
where $\Gamma_{i,j}^*$ is the data rate achieved at destination node $j$ when relay node $i$ is transmitting at power $P_{i,j}^*$. 

For our proposed auction, first, each relay $i$ declares its valuation, $\alpha_i$, to the BS. Then we construct a complete bipartite graph $G=(\mathbf{R},\mathbf{D},E)$. For each pair of nodes $(i,j)\in \mathbf{R} \times \mathbf{D}$, nodes $i$ and $j$ are connected by an edge whose weight is $U_{i,j}^*$ if $C_{i,j}(P_{i,j}^*)\leq C_{T}$, else $-\infty$. We denote the set of all possible maximal matchings as $\mathbf{M}$ and an individual matching by $m$. For every matching $m\subset E$, we define $w_m$ as the sum of weights of all the edges $(i,j)\in m$. For every matching $m$, we define a set $R_m$ which consists of all relay nodes that are in the neighbourhood of $\mathbf{D}$ under the matching $m$. If we denote $w_{max}=\max\limits_{m\in \mathbf{M}}w_m$ and $m_{max}=\argmax\limits_{m\in \mathbf{M}}w_m$, then we select the relay nodes in $R_{m_{max}}$ as the winners of the auction. Each relay node in $R_{m_{max}}$ is assigned to its neighbour in $\mathbf{D}$ under the maximal matching $m_{max}$. For every relay node $i$ we denote $w_{max}^{-i}=\max\limits_{m\in \mathbf{M}, i\notin R_m}w_m$. A relay node $i$ which is assigned to destination node $j$ is paid:
\begin{equation}
p_{i,j}=(w_{max}-w_{max}^{-i}+\alpha_i (P_{i,j}^*+P_{c,i})).
\label{eq:pay:relay:bsm}
\end{equation} 
The sequence of steps that implements our proposed auction is given in Fig.~\ref{algorithm:auction:BS:utility:maximization}.

\begin{figure}[!hbt]
\mbox{}\hrulefill \\
\begin{scriptsize}
\begin{algorithmic}[1]
\STATE{Construct a complete weighted bipartite graph $G=(\mathbf{R},\mathbf{D},E)$.}
\STATE{Define the weight of edge $(i,j)$ to be $U_{i,j}^*$ if $C_{i,j}(P_{i,j}^*)\leq C_{T}$, else $-\infty$.}
\STATE{Select a maximal matching $m_{max}$ such that $m_{max}=\argmax\limits_{m\in \mathbf{M}}w_m$, where $\mathbf{M}$ is the set of all maximal matchings and $w_m$ represents the sum of weights of all edges in the maximal matching $m$.}
\STATE{If $(i,j)\in m_{max}$ where $(i,j)\in\mathbf{R}\times\mathbf{D}$, then assign relay node $i$ to destination node $j$.}
\STATE{If relay node $i$ is assigned to destination node $j$, then it is paid $p_{i,j}=(w_{max}-w_{max}^{-i}+\alpha_i (P_{i,j}^*+P_{c,i}))$ and transmits at power $P_{i,j}^*$, else it is paid $0$ and relay node $i$ is not required to transmit any data.}
\end{algorithmic}
\end{scriptsize}
\mbox{}\hrulefill
\caption{\label{algorithm:auction:BS:utility:maximization} Auction to approximately maximize the BS's utility.}
\end{figure}

\begin{theorem}
The auction in Fig.~\ref{algorithm:auction:BS:utility:maximization} satisfies individual rationality and can be truthfully implemented.
\end{theorem}
\begin{IEEEproof}
Consider node $i$. Let $u_{i,k}$ denote the utility of relay node $i$ when it declares its valuation truthfully and is assigned to destination node $k$.  $k$ can be a pseudo user if relay node $i$ is not assigned to any destination node when it reveals its true valuation. Then $u_{i,k}$ is simply zero.  Let us assume that relay node $i$ manipulates its valuation and declares $\alpha_i'$ instead. This will change the weights of all matchings in the set $\mathbf{M}_i^c=\{m\in \mathbf{M}: (i,l)\in m \mbox{ for some } l\in \mathbf{D}\}$. Let the new weight of the matching $\bar{m}\in \mathbf{M}_i^c$ be $w_{\bar{m}}'$. If some matching $m\in \mathbf{M}_i^c$ is the matching with maximum weight $w_m'$, then relay node $i$ is assigned to a destination node. Otherwise relay node $i$ is not assigned to any destination node in which case its utility is $0$.  Assume that some $m\in \mathbf{M}_{i}^c$ is the matching with the maximum weight $w_m'$ and that relay node $i$ is assigned to destination node $j$ in matching $m$. Suppose when relay node $i$ declares $\alpha_i'$, it is assigned transmit power $P_{i,j}'$, and $\Gamma_{i,j}'$ is the data rate achieved at destination node $j$ when relay node $i$ transmits at power $P_{i,j}'$; also, let us denote the new weight of the edge $(i,j)$ as $U_{i,j}'$. We then have the following equality: $w_m'-U_{i,j}'=w_m-U_{i,j}^*$. Now by \eqref{eq:utility1}, \eqref{eq:energy} and \eqref{eq:pay:relay:bsm}, the utility of relay node $i$ is given by:
\begin{align*}
u_{i,j}'=&w_{m}'-w_{max}^{-i}+\alpha_i' (P_{i,j}'+P_{c,i})-\alpha_i (P_{i,j}'+P_{c,i})\\
=&w_m-U_{i,j}^*+U_{i,j}'-w_{max}^{-i}+\alpha_i' (P_{i,j}'+P_{c,i})\\&-\alpha_i (P_{i,j}'+P_{c,i}) \ \mbox{ (since } w'_{m}-U'_{i,j}=w_{m}-U^*_{i,j}) \\
=&w_m-w_{max}^{-i}+a\Gamma_{i,j}'-\alpha_i'(P_{i,j}'+P_{c,i})-C_{i,j}(P_{i,j}')\\
&-a\Gamma_{i,j}^*+\alpha_i (P_{i,j}^*+P_{c,i})+ C_{i,j}(P_{i,j}^*)\\&+\alpha_i'(P_{i,j}'+P_{c,i})-\alpha_i (P_{i,j}'+P_{c,i})\hspace{8mm}\mbox{(by \eqref{eq:BS:utility:final})}\\
=&w_m-w_{max}^{-i}+a\Gamma_{i,j}'-\alpha_i (P_{i,j}'+P_{c,i})-C_{i,j}(P_{i,j}')\\
&-(a\Gamma_{i,j}^*-\alpha_i (P_{i,j}^*+P_{c,i})-C_{i,j}(P_{i,j}^*))\\
\leq& w_m-w_{max}^{-i}
\end{align*}
The inequality holds because when relay node $i$ declares its valuation truthfully, $P_{i,j}^*$ maximizes $U_{i,j}$ (see \eqref{eq:U:maximizer}). If node $i$ is not selected when it declares its valuation truthfully, then we have $w_m-w_{max}^{-i}=w_m-w_{max}\leq 0$ and if node $i$ is selected when it declares its valuation truthfully, then  $w_{m}-w_{max}^{-i}\leq w_{max}-w_{max}^{-i}=u_{i,k}$. This proves that the above auction can be truthfully implemented. Also, since from \eqref{eq:pay:relay:bsm}, the utility of relay node $i$ assigned to a destination node $j$ is $w_{max}-w_{max}^{-i}\geq 0$, the proposed auction satisfies the individual rationality property. The result follows.
\end{IEEEproof}
\begin{proposition}
\label{prop:time:BS:utility}
The time complexity of the auction in Fig.~\ref{algorithm:auction:BS:utility:maximization} is $O(D\mathcal{H})$.
\end{proposition}
\begin{IEEEproof}
Relays are assigned to destination nodes by finding the maximum weighted maximal matching of the bipartite graph $G=(\mathbf{R},\mathbf{D},E)$ (see step 3 in Fig.~\ref{algorithm:auction:BS:utility:maximization}). This can be done using the Hungarian algorithm~\cite{RF:Hungarian} and the time complexity of this operation is $O(\mathcal{H})$.

Next, to compute the payment to relay node $i$ assigned to destination $j$ (see \eqref{eq:pay:relay:bsm}), $w_{max}^{-i}$ needs to be found. This can be  computed by finding the maximum weighted maximal matching of the bipartite graph $G^{-i}=(V\setminus i,E^{-\{i\}})$, where $E^{-\{i\}}=\left\{e\in E:e\neq (i,k), \forall k\in \mathbf{D}\right\}$, using the Hungarian algorithm. So the time complexity of computing the payment made to each of the $D$ relay nodes assigned to destination nodes is $O\left(D\mathcal{H}\right)$. The result follows.
\end{IEEEproof}

\subsection{Expressions for the Transmission Power of a Relay}
\label{sec:appendix:2}
In this subsection, we provide expressions for the transmission power of a relay node under various relaying schemes for the constant data rate scenario and approximate BS utility maximization scenario. Throughout this subsection, we assume for tractability that the interference cost to the BS (see \eqref{eq:bs:utility}) is $C_{i,j}(P_{i,j})=c_iP_{i,j}$. 

Let $\gamma_{i,j}=\frac{G_{i,j}}{N_{i,j}+I_{i,j}}$ and $\gamma_{s,j}=\frac{G_{s,j}}{N_{s,j}+I_{s,j}}$. Let $P_m$ denote  the maximum transmission power of a relay node. $P_{i,j}$ is the power at which relay node $i$ is required to transmit under the constant data rate scenario if it is assigned to destination node $j$, which requests a data rate of $\Gamma_{j}$, and $P_{i,j}^*$ is the power at which relay node $i$ is required to transmit if it is assigned to destination node $j$ to approximately maximize the BS's utility.

\subsubsection{Normal Relaying}
For the constant data rate scenario, when relay node $i$ is assigned to destination node $j$, which requests a data rate of $\Gamma_{j}$, by \eqref{eq:data:rate:normal}, the data rate $\Gamma_{j}$ in terms of the transmit power $P_{i,j}$ is given by:
\begin{equation*}
\Gamma_{j}=\frac{W}{2}\log_2(1+P_{i,j}\gamma_{i,j}).
\end{equation*}
From the above equation, the transmission power $P_{i,j}$ required by relay node $i$ is:
\begin{equation*}
P_{i,j}=\frac{4^{\frac{\Gamma_{j}}{W}-1}}{\gamma_{i,j}}.
\end{equation*}

Next, from \eqref{eq:max:BS:utility:2}, at the transmission power $P_{i,j}^*$ which approximately maximizes the BS's utility, we have:
\begin{equation}
\label{eq:BS:diff}
\frac{\partial U_{i,j}}{\partial P_{i,j}}=a\frac{\partial \Gamma_{i,j}}{\partial P_{i,j}}-\alpha_i-c_i=0.
\end{equation}
Substituting \eqref{eq:data:rate:normal} in the above equation and solving for $P_{i,j}^*$~\footnote{When $P_{i,j}^*<0$, we set $P_{i,j}^*=0$ and when $P_{i,j}^*>P_m$, we set $P_{i,j}^*=P_m$. This process is followed for all the relaying schemes. }, we get:
\begin{equation*}
P_{i,j}^*=\frac{aW}{2\ln 2(\alpha_i+c_i)}-\frac{1}{\gamma_{i,j}}.
\end{equation*}

\subsubsection{Amplify-and-Forward}
For the constant data rate scenario, when relay node $i$ is assigned to destination node $j$, we obtain the transmission power required by relay node $i$ as in the normal relaying scheme. Putting $\Gamma_{i,j} = \Gamma_{j}$ in \eqref{eq:data:rate:amplify:forward} and solving, we get:
\begin{equation*}
P_{i,j}=\frac{\left(4^{\frac{\Gamma_{j}}{W}}-1-SINR_{s,j}\right)\left(1+SINR_{s,i}\right)}{\left(1+SINR_{s,i}+SINR_{s,j}-4^{\frac{\Gamma_{j}}{W}}\right)\gamma_{i,j}}.
\end{equation*}

In the approximately maximizing the BS utility scenario,  we follow the same procedure as used for the normal relaying scheme. We find roots of \eqref{eq:BS:diff} by substituting \eqref{eq:data:rate:amplify:forward} for $\Gamma_{i,j}$ and as a result we get a quadratic equation. It can be easily seen that one root is always negative and hence we take the larger root of the quadratic equation, which is as follows, as the transmission power~\footnote{The transmission power is $0$ if both roots are negative.} of relay node $i$:
\begin{eqnarray*}
&&(1+SINR_{s,i} + SINR_{s,j}) P_{i,j}^2\gamma_{i,j}^2 \\ 
&&+{}{}(1+SINR_{s,i})(2+2SINR_{s,j}+SINR_{s,i}) P_{i,j}\gamma_{i,j} \\
&&+ {}{}(1+SINR_{s,j})(1+SINR_{s,i})^2 \\
&&= aW\dfrac{ SINR_{s,i} (1+SINR_{s,i})}{\ln4 (\alpha_i+c_i)} \gamma_{i,j}.
\end{eqnarray*}

\subsubsection{Decode-and-Forward}
From \eqref{eq:data:rate:decode:forward}, there are two possible cases:

\emph{Case 1}:
If $\frac{W}{2}\log_2(1+SINR_{s,i})\geq\frac{W}{2}\log_2(1+SINR_{s,j}+SINR_{i,j})$ $\forall$ $0 < P_{i,j} \le P_m$, then we have:
\begin{equation}
\label{eq:decode:case:1}
\Gamma_{i,j} = \frac{W}{2}\log_2(1+SINR_{s,j}+SINR_{i,j}).
\end{equation}

In the constant data rate scenario, the transmission power at which relay node $i$ must transmit to destination node $j$ is given by:
\begin{equation*}
P_{i,j}=\frac{4^{\frac{\Gamma_{j}}{W}}-1-SINR_{s,j}}{\gamma_{i,j}},
\end{equation*}

In the approximately maximizing the BS utility scenario, the same procedure is followed as in the normal relaying case. We find the root of \eqref{eq:BS:diff} by substituting \eqref{eq:decode:case:1} for $\Gamma_{i,j}$. The transmission power required by relay node $i$ while transmitting to destination node $j$ for approximately maximizing the BS's utility is:
\begin{equation*}
P_{i,j}^*=\frac{aW}{\ln4 (\alpha_i+c_i)}-\frac{1+SINR_{s,j}}{\gamma_{i,j}}
\end{equation*}

\emph{Case 2}: In Case 2, $\exists P_0$ such that $\frac{W}{2}\log_2(1+SINR_{s,i}) > \frac{W}{2}\log_2(1+SINR_{s,j}+SINR_{i,j})$ $\forall$ $0 < P_{i,j} < P_o$ and $\frac{W}{2}\log_2(1+SINR_{s,i}) \le \frac{W}{2}\log_2(1+SINR_{s,j}+SINR_{i,j})$ $\forall$ $P_o \le P_{i,j} \le P_m$. 

In the constant data rate scenario, in the case when $0 < P_{i,j} < P_o$, we have $\frac{W}{2}\log_2\left(1+SINR_{s,i}\right) > \frac{W}{2}\log_2\left(1+SINR_{s,j}+SINR_{i,j}\right)$; so:
\begin{equation*}
\Gamma_{i,j}=\frac{W}{2}\log_2\left(1+SINR_{s,j}+SINR_{i,j}\right)
\end{equation*}
The expression for $P_{i,j}$ in this case is the same as in Case 1.
In the case when $P_o \le P_{i,j} \le P_m$, we have $\frac{W}{2}\log_2(1+SINR_{s,i}) \le \frac{W}{2}\log_2(1+SINR_{s,j}+SINR_{i,j})$; so:
\begin{equation*}
\Gamma_{i,j}=\frac{W}{2}\log_2(1+SINR_{s,i}).
\end{equation*}
Since the data rate is independent of $P_{i,j}$, we set $P_{i,j}=P_o$ as this will minimize the interference cost to the BS (see \eqref{eq:bs:utility}).

In the approximately maximizing the BS utility scenario,  in Case 2, we find the maximum contribution to the utility of the BS across the two cases: $0<P_{i,j}^* < P_o$ and $P_o \leq P_{i,j}^*<P_m$. When we assume that $0<P_{i,j}^* < P_o$, we substitute $\Gamma_{i,j}=\frac{W}{2}\log_2\left(1+SINR_{s,j}+SINR_{i,j}\right)$ in \eqref{eq:BS:diff} and obtain the transmission power $P_{i,j}^*$ that maximizes \eqref{eq:max:BS:utility:2}. If $P_{i,j}^* \geq P_o$, then we set $P_{i,j}^*=P_o$ and if $P_{i,j}^*<0$, we set $P_{i,j}^*=0$. We find the corresponding contribution to the utility of the BS (see \eqref{eq:bs:utility}) which we denote by $U_1^*$.  Next, we repeat the process assuming that  $P_o \leq P_{i,j}^*<P_m$. In this case we substitute $\frac{W}{2}\log_2\left(1+SINR_{s,i}\right)$ as the expression for data rate in \eqref{eq:BS:diff}. It can be seen that $P_{i,j}^*=P_o$. We calculate the corresponding contribution to the BS utility (see \eqref{eq:bs:utility}), which we denote by $U_2^*$. If $U_1^*>U_2^*$, then the expression for $P_{i,j}^*$ is similar to the one obtained in Case 1, else $P_{i,j}^*=P_o$.

\subsubsection{Selection Relaying}
From \eqref{eq:data:rate:selection:relaying}, the capacity of the selection relaying protocol if relay node $i$ is assigned to destination node $j$ is given by:
\begin{equation*}
\Gamma_{i,j} = \frac{W}{2}\log_2(1+SINR_{s,j}+SINR_{i,j}).
\end{equation*}
The expressions for transmission powers are the same as those given in Case 1 of the decode-and-forward relaying scheme. Note that if $SINR_{i,j}<\zeta$, then relay $i$ is not assigned to destination node $j$.

\section{Comparison of Proposed auctions with VCG mechanism based auctions}
\label{sec:vcg}
We now compare the proposed auctions with those based on the VCG mechanism.

\subsection{Constant Power Case}
\label{SSC:comparison:proposed:vcg:const:power}
In this scenario, in the proposed auction, relay nodes are assigned to destination nodes such that the following expression is minimized (see Fig.~\ref{algorithm:auction:constant:power}):
\begin{equation}
\sum\limits_{j=1}^D\sum\limits_{i=1}^R\frac{\alpha_i(P+P_{c,i})}{\Gamma_{i,j}}y_{i,j},
\label{eq:proposed:constant:power}
\end{equation}
whereas in the VCG mechanism, relay nodes are assigned to destination nodes such that the following expression is minimized (see \eqref{eq:vcg:constant:power}):
\begin{equation}
\sum\limits_{j=1}^D\sum\limits_{i=1}^R\alpha_i(P+P_{c,i})y_{i,j},
\label{eq:vcg:constant:power:2}
\end{equation}
where $\sum\limits_{j=1}^{D}y_{i,j}\leq 1$ for all $i\in\mathbf{R}$ and $\sum\limits_{i=1}^{R}y_{i,j}=1$ for all $j\in \mathbf{D}$. From \eqref{eq:proposed:constant:power}, it can be seen that  
the data rates $\Gamma_{i,j}$ appear in the denominators of the terms in the quantity that the proposed auction seeks to minimize. Hence, the proposed auction tends to assign relay nodes to destination nodes in such a way that the achieved data rates at destination nodes are high. On the other hand, from \eqref{eq:vcg:constant:power:2}, it can be seen that the VCG mechanism based auction ignores the data rates $\Gamma_{i,j}$.  Also, by \eqref{eq:total:BS:utility} and \eqref{eq:bs:utility}, the utility
of the BS is an increasing function of the achieved data rates
of the destination nodes; hence, under the proposed auction, the utility of the BS tends to be high. Next, note that under the proposed auction, only those allocations of relays to destination nodes for which the costs ($C_{i,j}(\cdot)$) incurred due to interference caused by relays at the BS are sufficiently low can possibly be selected (see step 2 in Fig.~\ref{algorithm:auction:constant:power}); on the other hand, the VCG mechanism based auction ignores the interference cost. Due to the above reasons, the \emph{proposed auction outperforms the VCG mechanism based auction in terms of the data rates achieved by the destination nodes, the utility of the BS as well as the interference cost to the BS}; this is confirmed by the numerical results in Section~\ref{sec:numerical:results}.

\subsection{Constant Data Rate Case}
\label{SSC:comparison:proposed:vcg:const:data:rate}
In this case, the proposed auction assigns relay nodes to destination nodes such that the following expression is minimized (see Fig.~\ref{algorithm:auction:constant:data:rate}):
\begin{equation}
\sum\limits_{j=1}^D\sum\limits_{i=1}^R\frac{\alpha_i(P_{i,j}+P_{c,i})}{\Gamma_{j}}y_{i,j}. 
\label{eq:proposed:constant:data:rate}
\end{equation}
But by \eqref{eq:vcg:constant:data:rate},  under the VCG mechanism, relay nodes are assigned to destination nodes such that the following expression is minimized:
\begin{equation}
\sum\limits_{j=1}^D\sum\limits_{i=1}^R\alpha_i(P_{i,j}+P_{c,i})y_{i,j},
\label{eq:vcg:constant:data:rate:2}
\end{equation}
where $\sum\limits_{j=1}^{D}y_{i,j}\leq 1$ for all $i\in\mathbf{R}$ and $\sum\limits_{i=1}^{R}y_{i,j}=1$ for all $j\in \mathbf{D}$. It can be easily verified that the assignment of relay nodes to destination nodes under the proposed auction may differ from that under the VCG mechanism. Next, note that under the proposed auction, only those allocations of relays to destination nodes for which the costs ($C_{i,j}(\cdot)$) incurred  due to interference caused by relays at the BS are sufficiently low can possibly be selected (see step 2 in Fig.~\ref{algorithm:auction:constant:data:rate}); on the other hand, the VCG mechanism based auction ignores the interference cost. Due to this, the \emph{proposed auction outperforms the VCG mechanism based auction in terms of the interference cost to the BS}; this is confirmed by the numerical results in Section~\ref{sec:numerical:results}. 
 
\subsection{Selecting Power to Approximately Maximize the Utility of the BS}
As mentioned earlier, in this case, the VCG mechanism is not applicable since it does not specify how the transmission power should be set so as to approximately maximize the BS's utility.

\section{Numerical Results}
\label{sec:numerical:results}
In this section, we numerically evaluate the performance of the proposed auctions and compare it with that of the VCG mechanism based auctions.  

A hexagonal cell of radius 300 meters is considered with the relay nodes, destination nodes and the cellular users, which transmit on the uplink to the BS, placed using a uniform random distribution in the cell. At the beginning, we assign each channel to one destination node and one cellular user for uplink communication. If a relay node, say $i$, is assigned to a destination node, say $j$, during the auction, then $i$ transmits to $j$ on the channel assigned to $j$. For modelling the channel, we considered distance dependent path loss along with lognormal shadow fading. Also, the channel is assumed to undergo Rayleigh fading. The battery state, $b_i$, of each relay node $i\in \mathbf{R}$ is assumed to be distributed uniformly at random in the set $\{0.1,0.2,\ldots,1\}$. The value of $\alpha_{i}$ is taken to be the reciprocal of the battery state value $b_i$. Let the bandwidth of each channnel be $W$. Throughout, we take the number of destination nodes that request for relay services and the number of cellular users that transmit on the uplink in a given time slot to be $10$. We take the interference cost to be $C_{i,j}(P_{i,j})=c_i(\Gamma_{n_j}-\Gamma_{n_j}^i)$ (see Remark~\ref{RM:interference:cost}).

We evaluated the performance of the proposed and VCG mechanism based auctions under various relaying schemes in terms of the following metrics: data rates achieved by the destination nodes, utility of the BS and interference cost to the BS.  We repeated each experiment $100$ times and each time, independently, the channel gains and battery states were randomly chosen according to their distributions; the average values of the metrics over all the runs are depicted in the following plots.

The simulation parameters are given in Table~\ref{table:parameters}.
\begin{table}
\centering
\caption{Simulation Parameters}
\begin{tabular}{|c|c|}
\hline
Parameter & Value\\
\hline
Cell type & Hexagonal\\\hline
Cell radius & 300 m\\\hline
Propagation Model & \begin{tabular}{@{}c@{}}
Path loss with lognormal \\ shadow fading and Rayleigh \\fading\end{tabular}\\\hline
$a$ & 0.25 units per Mbps\\\hline
$c_i$ & 0.5a\\\hline 
Battery state & Uni. dist. in $\{0.1,0.2,\ldots,1\}$\\\hline
Channel Bandwidth  ($W$) & $10$ MHz\\\hline
Noise power & -174dBm/Hz\\\hline
Standard deviation for shadow fading & 8\\\hline
Path loss Exponent & 3.3 \\\hline
No. of destination nodes & 10\\\hline
$P_{m}$ &  24dBm\\\hline
$P_{s}$ &  4 W\\\hline
Interference Threshold ($C_T$) & $2.5$ units\\\hline
\end{tabular}
\label{table:parameters}
\end{table}

\subsection{Constant Power Case}
\begin{figure}[!hbt]
\centering
\includegraphics[scale=0.52]{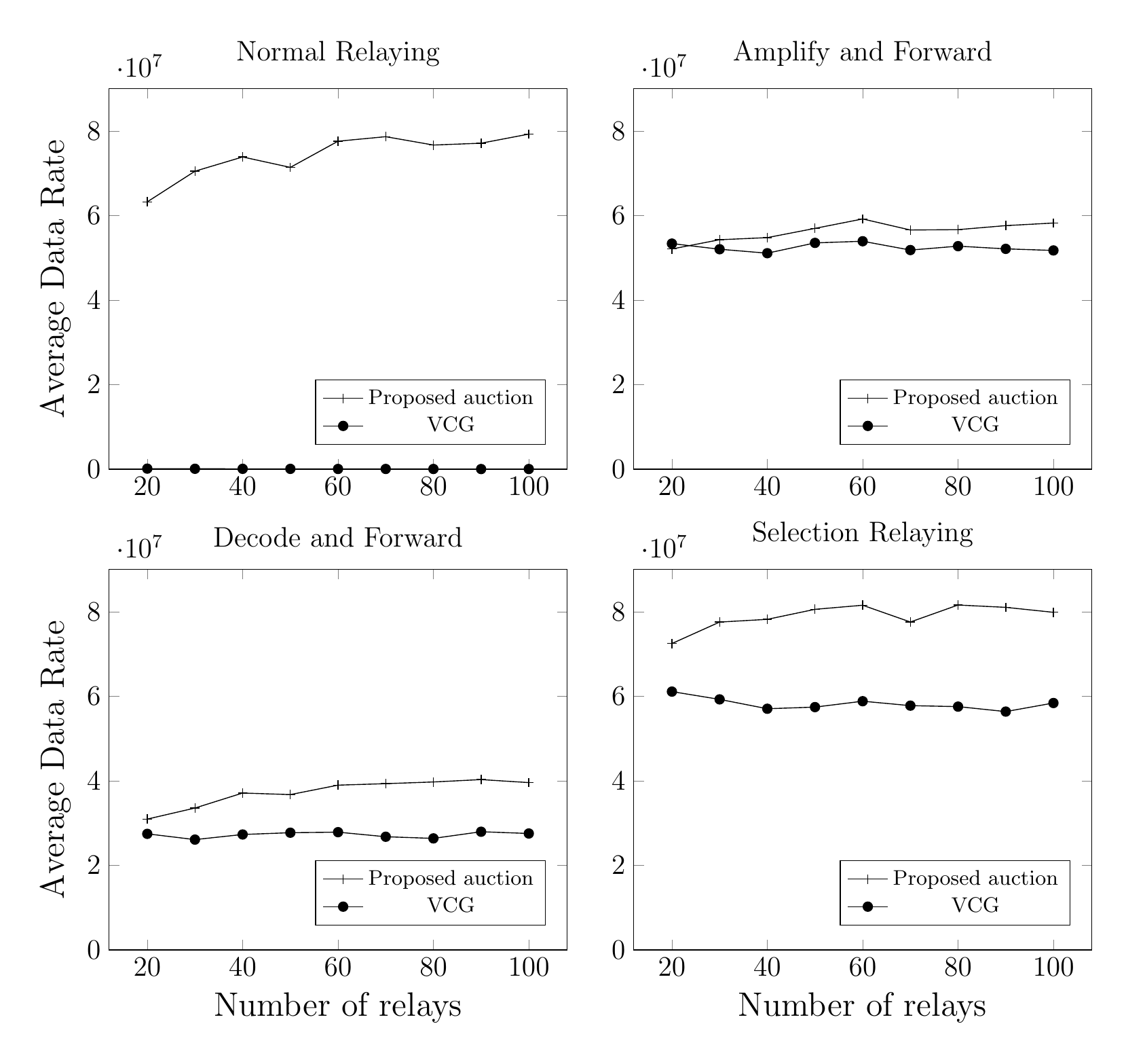}
\caption{The plots compare the average data rates achieved by destination nodes under the proposed auction with those under the VCG mechanism for various numbers of relay nodes and various relaying schemes for the constant power case.}
\label{fig:cpdr}
\end{figure}

\begin{figure}[!hbt]
\centering
\includegraphics[scale=0.52]{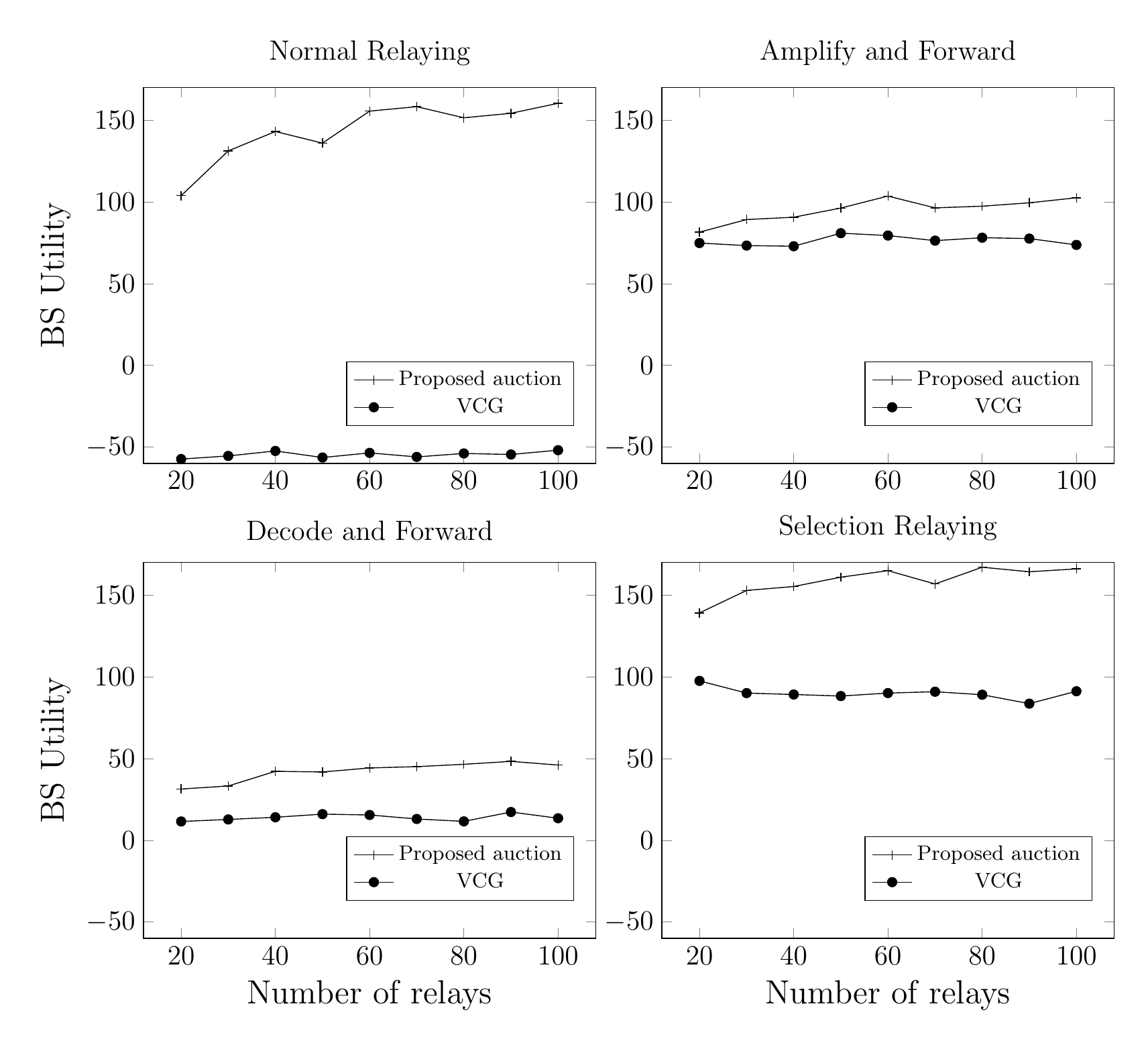}
\caption{The plots compare the average BS utilities under the proposed auction with those under the VCG mechanism for various numbers of relay nodes and various relaying schemes for the constant power case.}
\label{fig:cpbs}
\end{figure}

\begin{figure}[!hbt]
\centering
\includegraphics[scale=0.52]{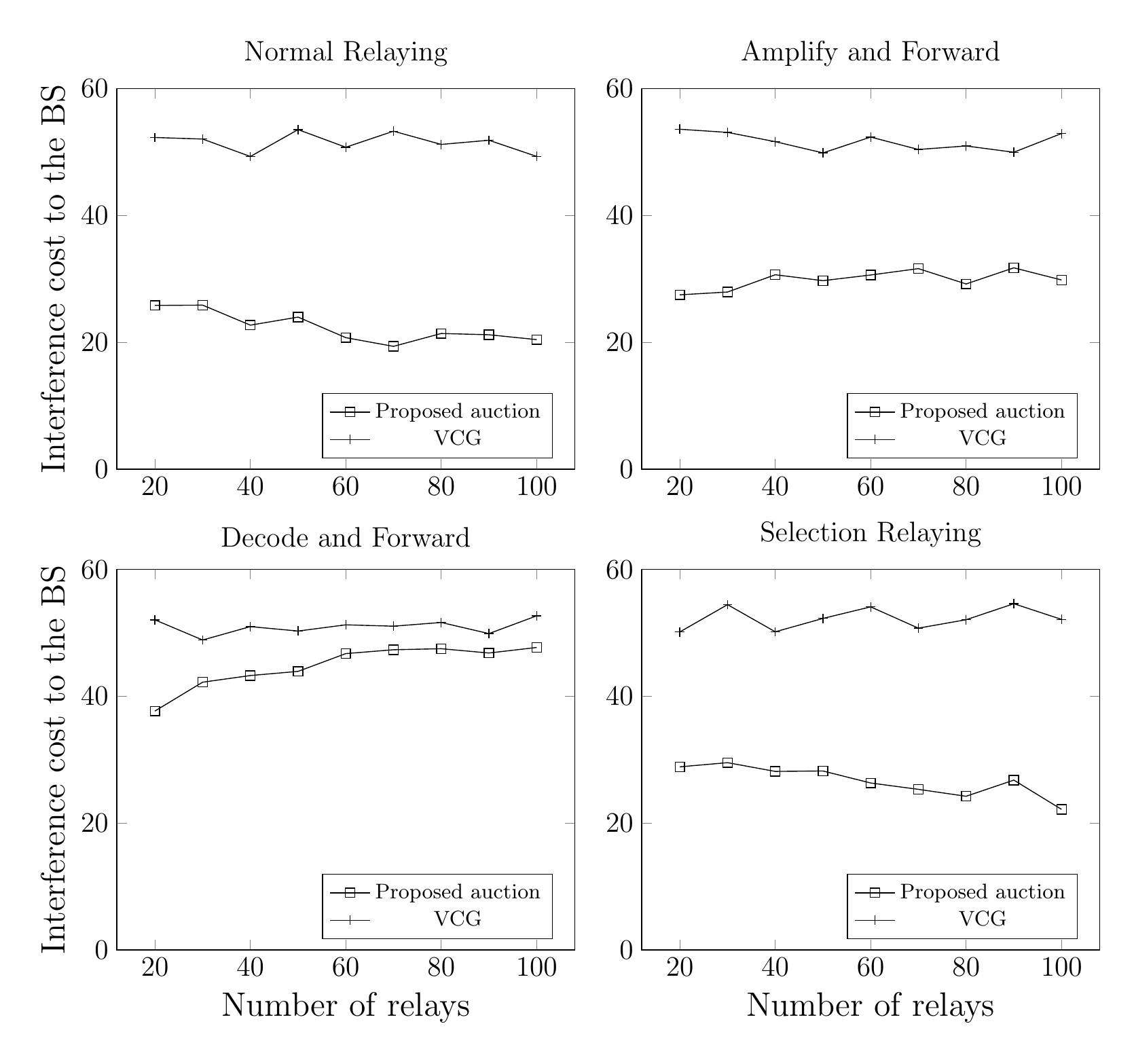}
\caption{The plots compare the average costs of interference to the BS under the proposed auction with those under the VCG mechanism for various numbers of relay nodes and various relaying schemes for the constant power case.}
\label{fig:cpin}
\end{figure}

The transmission power of each relay node was fixed at $P = 0.25$ W, while increasing the number of available relay nodes from $20$ to $100$. Fig.~\ref{fig:cpdr} shows the average data rate achieved per destination node versus the number of relay nodes for each of the four relaying schemes described in Section~\ref{subsection:relaying:schemes}. From the figure, it can be seen that \emph{for all four relaying schemes, the proposed auction outperforms the VCG mechanism based auction in terms of the achieved data rates}. Similarly,  Fig.~\ref{fig:cpbs} (respectively, Fig.~\ref{fig:cpin}) compares the average BS utility (respectively, the average interference cost to the BS $\left( \frac{1}{D}\sum_{i=1}^R \sum_{j = 1}^D C_{i,j}(P) y_{i,j} \right)$)  under the two auctions; it shows that \emph{for all four relaying schemes, the proposed auction outperforms the VCG mechanism based auction in terms of the average BS utility (respectively, average interference cost to the BS)}. The reasons for the trends in Figs.~\ref{fig:cpdr},~\ref{fig:cpbs} and~\ref{fig:cpin} are explained in Section~\ref{SSC:comparison:proposed:vcg:const:power}.

\subsection{Constant Data Rate Case}
\begin{figure}[!hbt]
\centering
\includegraphics[scale=0.52]{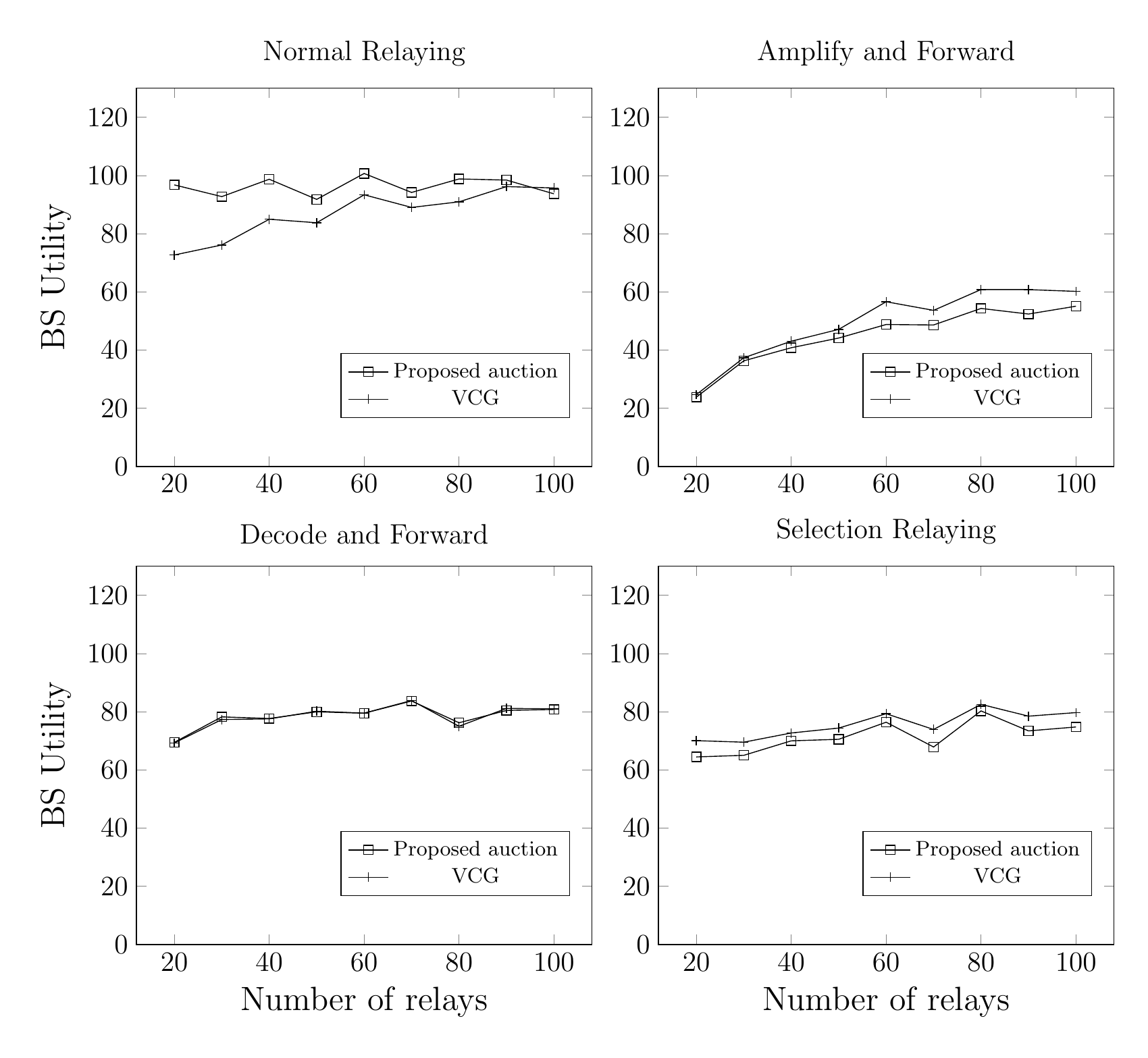}
\caption{The plots compare the average BS utilities under the proposed auction with those under the VCG mechanism for various numbers of relay nodes and for various relaying schemes for the constant data rate case.}
\label{fig:cdbs}
\end{figure}

\begin{figure}[!hbt]
\centering
\includegraphics[scale=0.52]{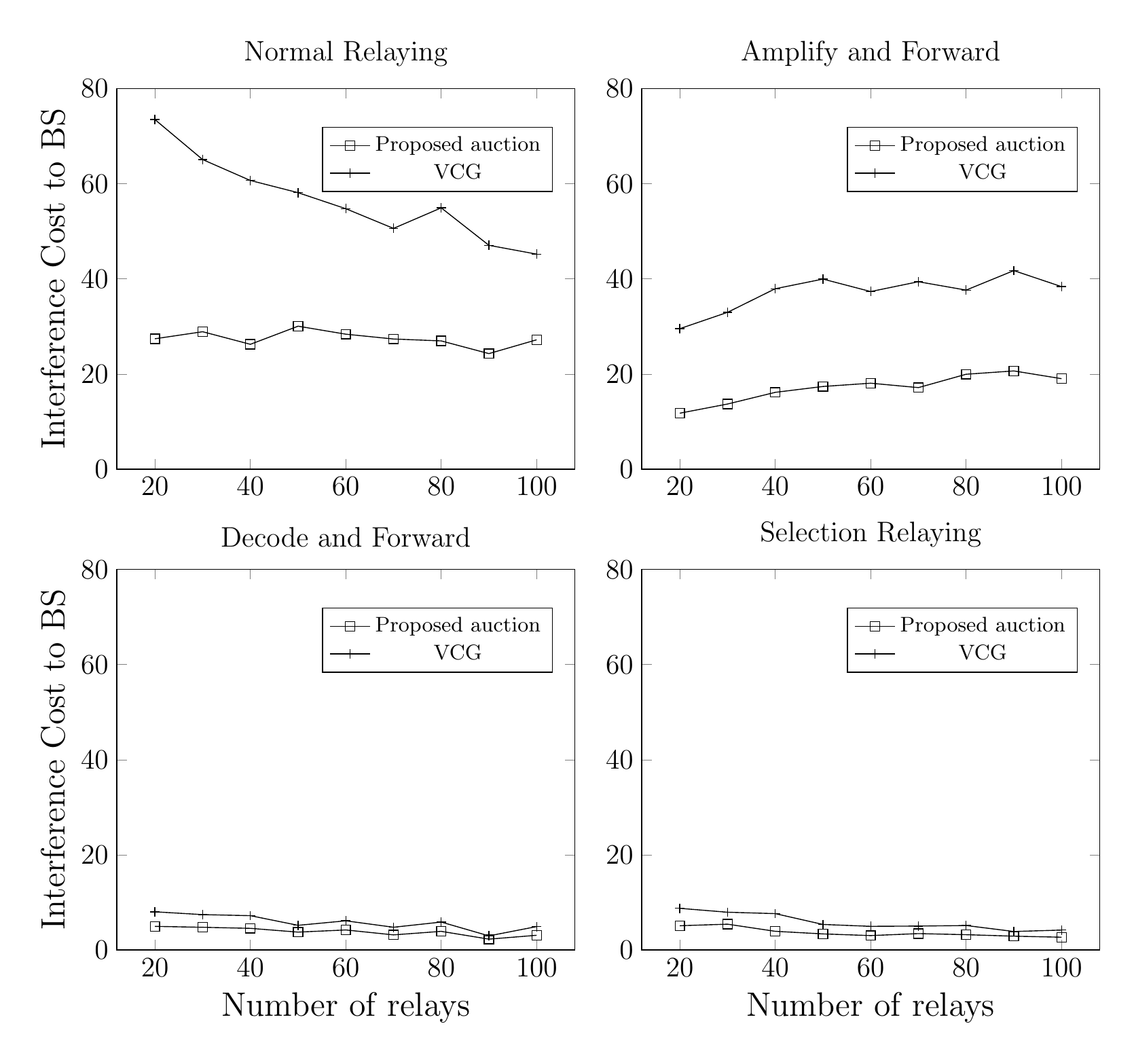}
\caption{The plots compare the average interference costs to the BS under the proposed auction with those under the VCG mechanism for various numbers of relay nodes and for various relaying schemes for the constant data rate case.}
\label{fig:cdin}
\end{figure}

Fig.~\ref{fig:cdbs}  compares the performance of the proposed auction with that of the VCG mechanism based auction in terms of the BS utility  for the constant data rate case and all four relaying schemes. The two auctions perform similarly in terms of the BS utility. Intuitively, this is because of \eqref{eq:total:BS:utility} and \eqref{eq:bs:utility} and  the fact that in the constant data rate case, the data rates at the destination nodes are the same ($\Gamma_j$) for both the proposed auction and the VCG mechanism based auction. Fig.~\ref{fig:cdin} compares the average interference cost to the BS under the two auctions for the four relaying schemes. The figure shows that \emph{the proposed auction outperforms the VCG mechanism based auction for all four relaying schemes in terms of the average interference cost to the BS}; the reason for this trend is explained in Section~\ref{SSC:comparison:proposed:vcg:const:data:rate}.

\subsection{Selection of Transmit Power to Approximately Maximize the BS's Utility}
\begin{figure}[!hbt]
\centering
\includegraphics[scale=0.52]{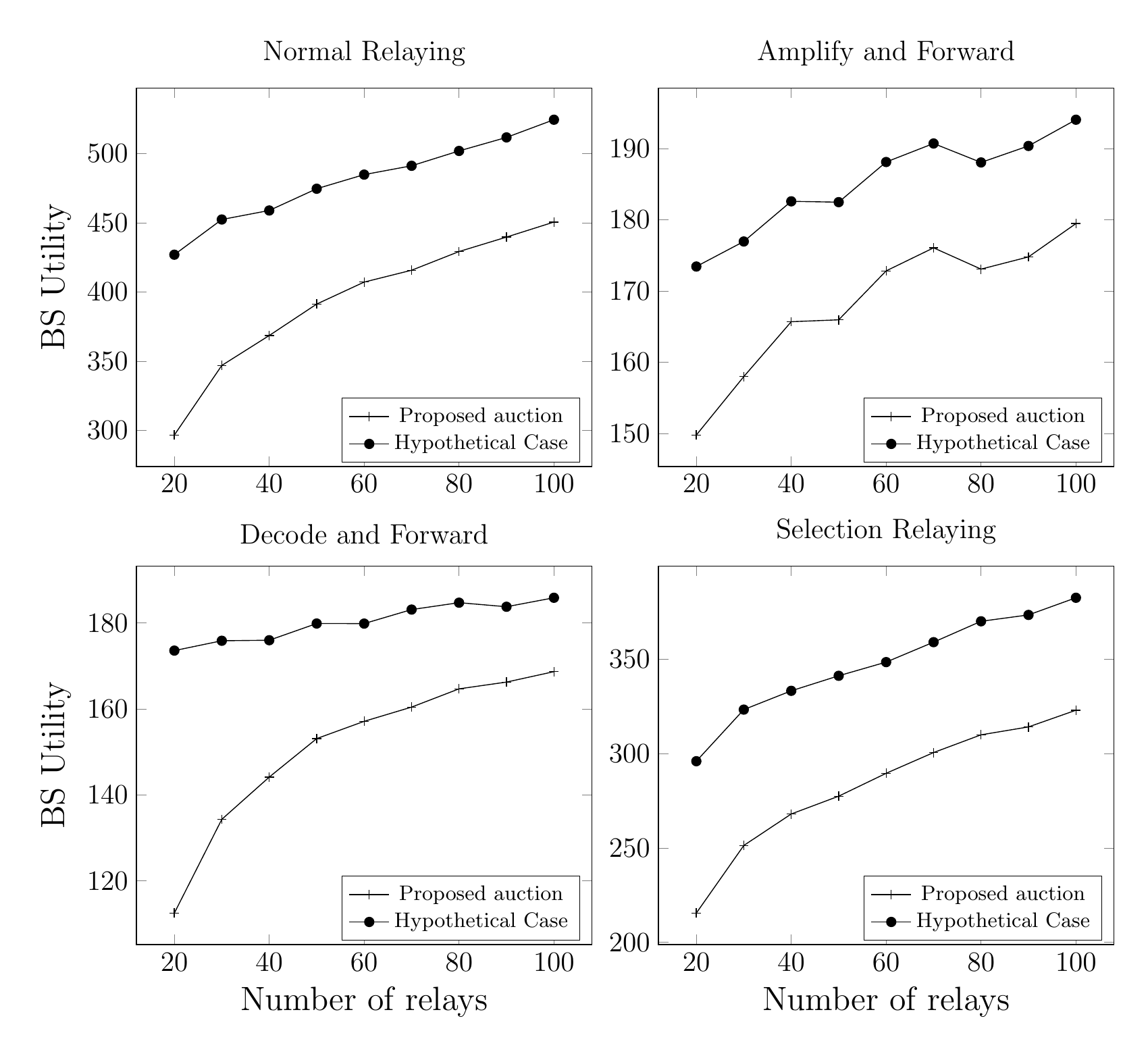}
\caption{In the BS utility maximisation case, for various relaying schemes and various numbers of available relay nodes, the above plots compare the average BS utilities under the proposed auction with those under the auction for the hypothetical case, which is similar to the former with the difference that every relay node is assumed to truthfully reveal its valuation and is paid only its incurred cost.}
\label{fig:bs:utility}
\end{figure}

We compare the BS's utility under the proposed auction with that under an auction for a hypothetical case, where nodes are assumed to always truthfully reveal their valuations $\alpha_i$. The latter auction makes the same assignment of relay nodes to destination nodes as in the proposed auction, but if relay node $i$ is assigned to destination node $j$, then the former is paid $\alpha_i(P_{i,j}^*+P_{c,i})$, i.e., each relay node is paid only its incurred cost. This is in contrast to the proposed auction where each selected relay node $i$ is paid an additional $w_{max}-w_{max}^{-i}$  (see \eqref{eq:pay:relay:bsm}). The plots in~ Fig.~\ref{fig:bs:utility} show that the BS's utility under the proposed auction is lower than that under the auction for the hypothetical case; this is because, when relay nodes may falsely declare their valuations (as in practice), they need to be paid more to incentivize them to truthfully declare their valuations. However, as the number of available relay nodes increases, the BS utilities under both the  auctions increase, but the difference between the utilities changes very little. Thus the proposed auction performs well even in large networks.

\section{Conclusions}
\label{SC:conclusions}
We considered a scenario in which some cellular users can relay data over D2D links to other cellular users with poor direct channel conditions from the BS. We addressed the problem of designing reverse auction mechanisms to assign a relay node to each destination node when there are multiple potential relay nodes and multiple destination nodes in each of the following three scenarios: 1) when relay nodes are allocated a fixed transmission power, 2)
when relay nodes are allocated the transmission powers required to achieve the
 data rates desired by the destination nodes, and 3) when the transmission powers of relay nodes are
selected so as to approximately maximize the BS's utility. We showed that in scenarios 1) and 2), auctions based on the widely used
VCG mechanism have several limitations, in particular, high interference cost and/ or low data rates achieved by destination nodes
and low BS utility; also, in scenario 3), the VCG mechanism is not applicable.
Hence, we proposed novel reverse auctions for relay
selection in each of the above three scenarios.
We proved that all the proposed reverse auctions can be
truthfully implemented as well as satisfy the individual rationality
property. Using numerical computations, we showed that in the fixed
transmission power scenario, our proposed auction significantly outperforms an
auction based on the widely used VCG mechanism in
terms of the data rates achieved by the destination nodes, the
utility of the BS and as well as the interference cost incurred to the BS; also, in the constant data rate case, our proposed auction outperforms the VCG mechanism based auction in terms of the interference cost to the BS. Our proposed auctions are applicable to a variety of relaying schemes such as Normal relaying, Decode-and-Forward relaying, Amplify-and-Forward relaying and Selection relaying.


\begin{thebibliography}{10}

\bibitem{RF:NCC:2017}
M.V.S. Aditya, P. Priyanka and G.S. Kasbekar,
\newblock ``Truthful Reverse Auction for Relay Selection, with High Data Rate and Base Station Utility, in D2D Networks",
\newblock In {\em Proc. of NCC}, Chennai, India, March 2017.

\bibitem{RF:keyword}
G. Aggarwal, A. Goel, and R. Motwani,
\newblock ``Truthful auctions for pricing search keywords",
\newblock In {\em Proc. of the 7th ACM conference on Electronic commerce}, New York, NY, USA, pp. 1-7, 2006.

\bibitem{Andrews20141065}
J. G. Andrews, S. Buzzi, W. Choi, S.V. Hanly, A. Lozano, A.C.K. Soong, J.C. Zhang,  
\newblock ``What will 5G be?'',
\newblock {\em IEEE Journal on Selected Areas in Communications}, Vol. 32, No. 6, pp. 1065-1082, Nov. 2014.

\bibitem{RF:D2D:survey}
A. Asadi, Q. Wang and V. Mancuso,
\newblock ``A Survey on Device-to-Device Communication in Cellular Networks",
\newblock In {\em IEEE CST}, Vol. 16, No. 4, pp. 1801-1819,  Fourthquarter 2014.


\bibitem{RF:bin:cao}
B. Cao, G. Feng, Y. Li and M. Daneshmand,
\newblock ``Auction-based Relay Assignment in Cooperative Communications",
\newblock In {\em Proc. of IEEE GLOBECOM}, Austin, TX, pp. 4496-4501, 2014.


\bibitem{RF:auction:offloading}
D. Chatzopoulos, M. Ahmadi, S. Kosta and P. Hui.
\newblock ``Have you asked your neighbors? A Hidden Market approach for device-to-device offloading",
\newblock In {\em Proc. of IEEE 17th WoWMoM},  Coimbra, pp. 1-9, June 2016.


\bibitem{RF:optimal:auction}
Y. Chen, S. He, F. Hou, Z. Shi and X. Chen,
\newblock ``Optimal user-centric relay assisted device-to-device communications: an auction approach",
\newblock In {\em IET Communications}, Vol. 9, No. 3, pp. 386-395, 2015.

\bibitem{RF:channel:estimation}
S. Coleri, M. Ergen, A. Puri and A. Bahai,
\newblock ``Channel estimation techniques based on pilot arrangement in OFDM systems",
\newblock In {\em IEEE Trans. on Broadcasting}, Vol. 48, No. 3, pp. 223-229, Sept. 2002. 

\bibitem{RF:combinatorial:optimization}
W. J. Cook, W. H. Cunningham, W. R. Pulleyblank and A. Schrijver,
\newblock {\em ``Combinatorial Optimization"},
\newblock Wiley-Interscience, 1997.


\bibitem{RF:doppler}
K. Doppler, M. Rinne, C. Wijting, C. B. Ribeiro and K. Hugl,
\newblock ``Device-to-device communication as an underlay to LTE-advanced networks",
\newblock In {\em IEEE Communications Magazine}, Vol. 47, no. 12, pp 42-49, Dec. 2009.

\bibitem{RF:data:transactions}
J. Du, E. Gelenbe, C. Jiang, Z. Han and Y. Ren,
\newblock ``Auction-based Data Transaction in Mobile Networks: Data Allocation Design and Performance Analysis",
\newblock In {\em IEEE TMC (Early Access)}, 2019.



\bibitem{RF:NUM}
J. Gao, L. Zhao and X. Shen,
\newblock ``Network Utility Maximization Based on an Incentive Mechanism for Truthful Reporting of Local Information",
\newblock In {\em IEEE TVT}, Vol. 67, No. 8, pp. 7523-7537, Aug. 2018.

\bibitem{RF:Gao}
L. Gao, G. Iosifidis, J. Huang and L. Tassiulas,
\newblock ``Hybrid data pricing for network-assisted user-provided connectivity",
\newblock In {\em Proc. of IEEE INFOCOM}, Toronto, ON, pp. 682-690, 2014.





\bibitem{RF:5G}
A. Ghosh, J. Zhang, J, Andrews, R. Muhamed
\newblock ``\emph{Fundamentals of LTE}",
\newblock Pearson Education, 2011.

\bibitem{RF:distributed:auction}
M. Hasan and E. Hossain,
\newblock ``Distributed Resource Allocation in D2D-Enabled Multi-tier Cellular Networks: An Auction Approach",
\newblock In {\em Proc. of IEEE ICC}. London, pp. 2949-2954, 2015.

\bibitem{RF:Christian:Isheden}
C. Isheden and G. P. Fettweis,
\newblock ``Energy-efficient multi-carrier link adaptation with sum rate-dependent circuit power",
\newblock In {\em Proc. of IEEE GLOBECOM}, pp. 1-6, Miami, FL, USA, 2010.


\bibitem{RF:online:auction}
T. Jing, F. Zhang, W. Cheng, Y. Hau and X. Cheng,
\newblock ``Online auction-based relay selection for cooperative communication in CR networks",
\newblock In {\em EURASIP Journal on Wireless Communications and Networking},  Feb. 2015.

\bibitem{RF:j:kim}
J. Kim and D. H. Cho,
\newblock ``A Joint Power and Subchannel Allocation Scheme Maximizing System Capacity in Indoor Dense Mobile Communication Systems",
\newblock In {\em IEEE TVT}, Vol. 59, No. 9, pp. 4340-4353, Nov. 2010.

\bibitem{RF:Hungarian}
H. W. Kuhn,
\newblock ``The Hungarian Method for the assignment problem",
\newblock In {\em Naval Research Logistics Quarterly}, Vol. 2, Issue 1-2, pp. 83-97, March 1955.

\bibitem{RF:Laneman}
J.N. Laneman, D.N.C. Tse and G.W. Wornell,
\newblock ``Cooperative diversity in wireless networks: Efficient protocols and outage behavior",
\newblock In {\em IEEE Trans. on Information Theory}, Vol. 50, No. 4, pp. 3062-3080, Dec. 2004.

\bibitem{RF:combinatorial:2}
D. Lehmann, L. I. O''Callaghan, and Y. Shoham,
\newblock ``Truth Revelation in Rapid, Approximately Efficient Combinatorial Auctions. Technical Report",
\newblock Stanford University, Stanford, CA, USA, 1999.


\bibitem{RF:Mi}
M. Li, W. Liao, X. Chen, J. Sun, X. Huang and P. Li,
\newblock ``Economic-Robust Transmission Opportunity Auction for D2D Communications in Cognitive Mesh Assisted Cellular Networks",
\newblock In {\em IEEE TMC}, Vol. 17, No. 8, pp. 1806-1819, Aug. 2018.

\bibitem{RF:YLi}
Y. Li, C. Liao, Y. Wang and C. Wang,
\newblock ``Energy-Efficient Optimal Relay Selection in Cooperative Cellular Networks Based on Double Auction",
\newblock In {\em IEEE TWC}, Vol. 14, No. 8, pp. 4093-4104, Aug. 2015.




\bibitem{RF:interference}
D. Lopez-Perez, A. Valcarce, G. de la Roche and J. Zhang,
\newblock ``OFDMA femtocells: A roadmap on interference avoidance",
\newblock In {\em IEEE Communications Magazine}, Vol. 47, No. 9, pp. 41-48, Sept. 2009.

\bibitem{RF:Ma}
Q. Ma, L. Gao, Y. Liu and J. Huang,
\newblock ``Economic Analysis of Crowdsourced Wireless Community Networks",
\newblock In {\em IEEE TMC}, Vol. 16, No. 7, pp. 1856-1869, July 2017.

\bibitem{RF:mascolell:microeconomic}
A. Mas-Colell, M.D. Whinston and J.R. Green,
\newblock {\em ``Microeconomic Theory"},
\newblock Oxford University Press, 1995.

\bibitem{RF:Mogensen}
P. Mogensen \emph{et al.},
\newblock ``LTE Capacity Compared to the Shannon bound",
\newblock In {\em Proc. of IEEE 65th. VTC-Spring}, Dublin, pp. 1234-1238, 2007.


\bibitem{RF:nambiar}
A. Nambiar and G. S. Kasbekar,
\newblock ``Complexity analysis and algorithms for the Inter Cell Interference Coordination with fixed transmit powers problem",
\newblock In {\em Proc. of IEEE COMSNETS}, pp. 1-8, Jan. 2015.


\bibitem{RF:UTRA}
RAN, TSG.
\newblock ``Requirements for further advancements for E-UTRA (LTE-Advanced)",
\newblock June 2008.

\bibitem{RF:algorithms}
R. Sedgewick and K. Wayne,
\newblock ``{\em Algorithms}",
\newblock 4th ed, Addison-Wesley Professional, 2011.


\bibitem{RF:y:wei}
Y. Wei and J.M. Cioffi,
\newblock ``Constant-Power Waterfilling: Performance-Bound and Low-Complexity Implementations",
\newblock In {\em IEEE Trans. on Communications}, Vol. 54, No. 1, pp. 23-28, Jan. 2006.


\bibitem{RF:Wen}
S. Wen et al.,
\newblock ``Achievable Transmission Capacity of Relay-Assisted Device-to-Device (D2D) Communication Underlay Cellular Networks",
\newblock In {\em Proc. of IEEE 78th VTC-Fall}, Las Vegas, NV, pp. 1-5, 2013.

\bibitem{RF:graph:theory:west}
D. B. West,
\newblock ``{\em Introduction to Graph Theory}",
\newblock 2nd ed. Princeton Hall, 2000.


\bibitem{RF:HERA}
D. Yang, X. Fang and G. Xue,
\newblock ``HERA: An Optimal Relay Assignment Scheme for Cooperative Networks",
\newblock In {\em IEEE JSAC}, Vol. 30, No. 2, pp. 245-253, Feb. 2012.


\bibitem{RF:WYong}
W. Yong, Y. Li, L. Chao and X. Yang,
\newblock ``Double-Auction-Based Optimal User Assignment for Multisource, Multirelay Cellular Networks",
\newblock In {\em IEEE TVT}, Vol. 64, No. 6, pp. 2627-2636, June 2015.


\bibitem{RF:chia-hao}
C.H. Yu et al.,
\newblock ``Resource Sharing Optimization for Device-to-Device Communication Underlaying Cellular Networks",
\newblock in {\em IEEE TWC}, vol. 10, No. 8, pp. 2752-2763, Aug. 2011.


\bibitem{RF:Zhang}
Y. Zhang, Y. He, J. Wang, Y. Kang, D. Liu, B. Li and Y. Liu,
\newblock ``Share Brings Benefits: Towards Maximizing Revenue for Crowdsourced Mobile Network Access",
\newblock In {\em Proc. of IEEE SECON}, San Diego, CA, pp. 1-9, 2017.

\bibitem{RF:spectrum:auction}
X. Zhou, S. Gandhi, S. Suri, and H. Zheng,
\newblock ``eBay in the Sky: strategy-proof wireless spectrum auctions",
\newblock In {\em Proc. of  MobiCom}, New York, NY, USA, pp. 2-13, 2008. 

\end{thebibliography}
\end{document}